\documentclass[twopage,11pt]{article}
\usepackage{amssymb}

\begin{document}

\title{Modification of the field theory and the dark matter problem}
\author{A. A. Kirillov \\
\emph{Institute for Applied Mathematics and Cybernetics} \\
\emph{10 Uljanova Str., Nizhny Novgorod, 603005, Russia}\\
} \maketitle

\begin{abstract}
We present an extension of the field theory onto the case in which the
topology of space can vary. We show that the nontrivial topology of space
displays itself in a multivalued nature of all observable fields and the
number of fields becomes an additional degree of freedom. In the limit in
which topology changes are suppressed, the number of fields is conserved and
the Modified Field Theory (MOFT) reduces to the standard field theory where
interaction constants undergo an additional renormalization and acquire a
dependence on spatial scales. This means that in MOFT particles lose their
point-like character and acquire a specific distribution in space, i.e. each
point source is surrounded with a halo which carries charges of all sorts.
From the dynamical standpoint such halos can be described as the presence of
a dark matter or fictitious particles.

When assuming that in the Planck stage of evolution of the Universe topology
changes do occur and that the early Universe is in thermodynamic
equilibrium, MOFT inevitably predicts the deviation of the law of gravity
from Newton's law in a certain range of scales $r_{\min }<r<r_{\max }$, in
which the gravitational potential shows essentially logarithmic $\sim \ln r$
(instead of $1/r$) behavior. In this range, the renormalized value of the
gravitational constant $G$ increases and at scales $r>r_{\max }$ acquires a
new constant value $G^{\prime }\sim Gr_{\max }/r_{\min }$.

We show that in MOFT fermions obey a generalized statistics and at scales $%
r>r_{\min }$ violate the Pauli principle (more than one fermion can occupy
the same quantum state). We also demonstrate that in this range the stable
equilibrium state corresponds to the fractal distribution of baryons and,
due to the presence of fictitious particles predicted by MOFT, this
distribution is consistent with observational limits on $\Delta T/T$. The
concept of fictitious particles is then used to explain the origin of the
diffuse component in the X-ray background and the origin of Higgs fields.
Thus, we show that MOFT allows to relate the rest mass spectrum of
elementary particles with cosmological parameters. Finally it is
demonstrated that in MOFT, in the range $r_{\min }<r<r_{\max }$ the Universe
acquires features of a two-dimensional space whose distribution in the
observed $3$- dimensional volume has an irregular character. This provides a
natural geometric explanation to the observed fractal distribution of
galaxies and the logarithmic behavior of the Newton's potential for a point
source. In conclusion we discuss some open problems in MOFT.
\end{abstract}





\pagestyle{myheadings} \thispagestyle{plain} \markboth{A. A.
Kirillov}{Modification of the field theory and the dark matter
problem} \setcounter{page}{1}

\section{Introduction}

The discrepancy between the luminous matter and the dynamic, or gravitating
mass was first identified in clusters of galaxies \cite{Zw}. Since then it
has widely been accepted that the leading contribution to the matter density
of the Universe comes from a specific non-baryonic form of matter (unseen,
or dark matter, see e.g. Refs. \cite{dm,dm2}). There are two basic arguments
in favor of dark matter and both depend essentially on the underlying
theory. First, if the mass distribution in a galaxy follows the brightness
the rotation curve is expected to show a Keplerian $r^{-1/2}$ law, while
measurements (e.g., see Ref. \cite{CF}) show a quite different behavior.
Namely, $V\left( r\right) \simeq V_{c}$ stays constant out to the visible
edge in most of galaxies. This means that the gravitating mass $M\left(
r\right)$ contained within a radius $r$ grows as $M\left( r\right)
=V_{c}^{2}r/G$ and a large fraction of the total mass of a galaxy has a
non-luminous dark form. This cannot be any normal form of matter, for the
normal matter can always be detected (e.g., intergalactic gas radiate X-rays
and is, therefore, seen). Note that there is a number of objects (normal
galactic matter - stars, dust, and gas - surrounding a galaxy) which give
direct evidence for a spherical halo of dark matter extending far beyond the
optical disk of a galaxy \cite{R84}. Analysis of the mass-to-light ratio of
galaxies, groups, and clusters (e.g., see Ref. \cite{B94}) shows that while
the $M/L$ ratio of galaxies increases with scale, it flattens and remains
approximately constant for groups and clusters. This means that if there is
an additional amount of dark matter in clusters (different from that of
galactic halos), it should contribute only to the homogeneous background.

The second argument is that the Universe has a rather developed structure
(galaxies, clusters, super-clusters). Metric potential fluctuations are
measured directly by $\Delta T/T$ in the microwave background \cite{Bar},
and observational values are at least two orders less than it is required by
the baryon-dominated Universe. Moreover, it was shown (e.g., see Refs. \cite%
{Ruf,P87,CP} and for a more recent discussions Refs. \cite{LMP,R99,FrD})
that the observed galaxy distribution exhibits a fractal behavior with
dimension $D\approx 2$ which seems to show no evidence of cross-over to
homogeneity. Such a picture is in a conflict with the Friedman model unless
a substantial amount of dark matter is present to restore the homogeneity of
the Universe. Note that this does not remove the conflict with the observed
values of $\Delta T/T$, for at the moment of recombination the density of
baryons and the CMB temperature are related as $n_{b}\sim T^{3}$ and,
therefore, the fractal distribution of baryons must leave a direct imprint
in the CMB temperature \cite{K02}.

Apart from some phenomenological properties of the dark matter (e.g., it
starts to show up in galactic halos, it is non-baryonic, cold, etc.) nothing
is known of its nature. Particle physics suggests various hypothetical
candidates for dark matter. We, however, still do not observe such particles
in direct laboratory experiments, while the dark matter displays itself by
the gravitational interaction only. Besides, there appears another puzzle
that the distribution of the luminous matter traces rather rigidly
perturbations in the density of the dark matter (the so-called biased galaxy
formation). While such a behavior may be acceptable at the non-linear or
quasi-linear stages of the development of perturbations, it looks quite
strange and cannot be explained (at least by the presence of hypothetical
particles) for the linear stage (e.g., at scales of superclusters where
perturbations in the total density are still small $\delta \rho _{tot}/\rho
_{tot}\ll 1$). All these facts suggest to try, as an alternative to the dark
matter hypothesis, the possibility to interpret the observed discrepancy
between luminous and gravitational masses as a violation of the law of
gravity.

The best known attempt of such kind is represented by a phenomenological
algorithm by Milgrom \cite{mil}, the so-called MOND (Modified Newtonian
Dynamics). This algorithm suggests replacing the Newton's law of gravity, in
the low acceleration limit $g\ll a_{0}$, with $g_{\mathrm{MOND}}\sim \sqrt{
ga_{0}}$, where $g$ is the gravitational acceleration and $a_{0}$ is a
fundamental acceleration constant $a_{0}\sim 2\times 10^{-8}cm/\,s^{2}$).
This, by construction, accounts for the two observational facts: the flat
rotation curves of galaxies (for remote stars from the center of a galaxy
MOND gives $g_{\mathrm{MOND}}=\sqrt{GM_{gal}a_{0}}/r=V_{c}^{2}/r$) and the
Tully-Fisher relation $L_{gal}\propto V_{c}^{4}$ which gives $M_{gal}
\propto L_{gal}\propto V_{c}^{4}$ (where $L_{gal}$, $M_{gal}$, and $V_{c}$
are the galaxy's luminosity, mass, and rotation velocity respectively). MOND
was shown to be successful in explaining properties of galaxies and clusters
of galaxies \cite{Sand,Sand2}, and different aspects of it still attract
attention, e.g., see Refs.\cite{Mc,Mul,L,L2,Mil} and references therein (see
also criticism in Ref. \cite{SW}).

However, in the present form MOND is not widely accepted, for the lack, in
the first place, of a clear theoretical motivation of such a nonlinear
behavior from particle physics standpoint (at low accelerations the force $%
F\propto \sqrt{M}$). Indeed, in particle physics the Newton's law of gravity
reflects merely the fact that gravitons are massless particles. There exist
processes which are able to violate the standard Newton's law, e.g., vacuum
polarization effects are known to produce corrections to the Coulomb
potential \cite{LL} and analogous corrections are expected to exist for the
Newton's potential (e.g., see Ref. \cite{ob,ob1} where some observational
limits on the scale-dependence of the gravitational constant were
considered). Such corrections, however, are essential at scales smaller than
the Planck length $\ell \lesssim \ell _{pl}\sim 10^{-33}cm$ where quantum
gravity effects are believed to be important and this, obviously, differs
from the typical scale of a galaxy ($\ell \propto 30-50Kpc$ $\propto
10^{23}cm$) where we would expect the gravity law to violate.

It turned out, however, that quantum gravity effects are, nevertheless,
capable of violating the Newton's law at large spatial distances, if
possible changes in the topology of space are taken into account (e.g., see
Ref. \cite{KT02}). On the classical level topology changes are known to be
forbidden. Such changes were, however, to occur during the quantum period of
the evolution of the Universe and, therefore, a relic of topology
transformations may still survive \cite{K99}. As it was shown in Ref. \cite%
{KT02}, the nontrivial topological structure of space displays itself in a
renormalization of all interaction constants, which, in general, depends on
spatial scales. This does mean the violation of usual interaction laws in
some range of scales. Specifically, if on the quantum stage the Universe is
assumed to be thermalized with a very high temperature, then one can show
\cite{KT02} that Newton's potential has to transform to essentially
logarithmic one ($\sim \ln r$ instead of $1/r$) in a certain range of scales
$r_{\min }<r<r_{\max }$. If we identify $r_{\min}$ with the characteristic
size of a galaxy, this amplification of the gravity force will produce flat
rotation curves without the dark matter hypothesis. We stress that the
switch to the logarithmic potential is a rather stringent prediction which
does not depend on details of the complete theory (and, in particular, on
quantum gravity). In other words, if we believe in topology changes in the
very early Universe, we should be very surprised if we would not see a
discrepancy between the luminous and dynamical masses.

In fact, the same kind of conclusions holds true for the Coulomb potential
and for all other interactions. In this sense we should observe not only
dark matter, but dark charges of all sorts as well (and it is very probable
that we do).

To illustrate the way how nontrivial topology of space causes a
renormalization of charge and mass values we consider a toy example. Let $q$
be the electric charge of a particle in a flat space which includes also a
number of handles, and $L$ be the characteristic size of the handles (the
distance between end points). In the Coulomb's field of the particle every
handle works as a dipole with a moment $d=\delta qL$, for it seizes some
fraction of lines of electric force of the particle. It is clear that the
farthest end of a handle acquires a charge $\delta q$ of the same sign as
that of the particle, while the nearest end acquires the charge of the
opposite sign $-\delta q$. The value $\delta q$ depends on the distance
between the closest end of the handle and the particle and on the
characteristic size of the throat of the handle as well (we note that the
value $\delta q$ is always less than $q$).

Consider now a ball of a radius $r$ with the particle in the center. In
general this ball includes a number of ends of the handles, and while $r\ll
L $ every handle gets into the ball by one end only. Suppose that there is
only one such a handle. Then, the observed value of the charge within the
ball of radius $r\ll L$ will be diminished by the value $\delta q$, while
for $r\gg L$ the ball will include both ends of the handle and the total
value of the charge restores. Hence, $\delta q$ can be interpreted as a dark
charge residing between the ends of the handle. In other words every handle
transports some fraction of the charge of the particle over a distance of
the order $L$, so a distributed set of handles will create a specific halo
of dark charge around every point source. Clearly, we have exactly the same
picture for the gravitational potential of the particle, i.e. the handles
create the halo of dark mass as well. Properties of the halo depend on the
distribution of handles, i.e. on specific properties of the topological
structure of space.

Apart from the changes in the gravity law, the qualitative picture shown
above gives at least two direct predictions. First, it can explain the
origin of the diffuse component of the X-ray background, for the presence of
dark charge extends considerably the characteristic size of a region
occupied with plasma (the hot X-ray emitting gas). And secondly, we should
expect the existence of fluctuations in the fine structure constant (i.e.,
in the value of the electric charge). Indeed, if we consider a ball of a
radius $r\ll L$ in a moving (with respect to handles) frame, then every
handle crossing the ball will cause some variation in the total charge $%
Q\left( t\right) =q-\delta q\left( t\right) $ (where $\delta q\left(
t\right) =0$ when the handle is outside the ball). This problem, however,
requires a more rigorous consideration.

The quantitative description of the situation is somewhat different from the
simple picture above. It is based on the suggestion of Ref. \cite{K99} that
the nontrivial topology of space should display itself in the multivalued
nature of all observable fields, i.e. the number of fields should be a
dynamical variable. The argument is that in the case of general position an
arbitrary quantum state mixes different topologies of space. From the other
side, any measurement of such a state should be carried out by a detector
which obeys classical laws and, therefore, the detector introduces a
background space of a particular topology in terms of which quantum states
should be described. It is important that on the classical level the
topology is always defined and does not change (i.e., if the Universe was in
a particular initial quantum state which mixes different topologies of
space, it should eventually remain in the mixed topology state). This means
that the topology of space must not be a direct observable, and the only
chance to keep the information on the topology is to allow all the fields
(which are specified on the background space) to be multivalued. It turns
out that a very good compatibility of the multivalued field theory with the
conventional one is achieved when the variable number of fields is
introduced in the momentum representation, i.e. for Fourier transforms. In
other words the topology has to be fixed for the space of momenta, while in
the coordinate space it is not defined at all.

The Modified Field Theory (MOFT) which treats multivalued fields was
developed in Refs. \cite{K99,KT02,K03}. Despite MOFT is a self-consistent
theory from particle physics standpoint, and basic features and principles
of constructing the theory are rather transparent, it is still not
sufficiently elaborated. At present we get only particular fragments which
we discuss in the present paper, while a satisfactory model of topology
transformations is still missing. However the existing fragments seem to be
promising and rather impressive. First of all, MOFT provides a natural
explanation to the dark matter phenomenon \cite{KT02}. Secondly, the
theoretical scheme of MOFT reconciles the observed developed structure of
the Universe (in particular, the fractal distribution of luminous matter)
with the homogeneity of the Universe and with the observational limits on $%
\Delta T/T$ in the microwave background \cite{K02}. MOFT also provides a
number of new predictions which can be used to verify the theory. In
particular, dark charge predicted by MOFT may be responsible for the
formation of the diffuse component in the X-ray background, whose nature
seems to be analogous to that of CMB (and very likely for the formation of
galactic magnetic fields). Besides, as it is shown in the present paper,
MOFT allows to explain the origin of the observed rest mass spectrum of
elementary particles and to relate it to cosmological parameters.

\section{One-dimensional crystal. Basic ideas of MOFT}

We start our consideration with the most simple illustrative example.
Consider a one-dimensional crystal. Positions of an atom in the crystal can
be described by a single coordinate $x$. In quantum mechanics states of an
atom will be described by a wave function $\psi \left( x\right) $. When
considering systems with a variable number of atoms the wave function
becomes an operator $\widehat{\psi}\left( x\right)$ which annihilates (and $%
\widehat{\psi}^{+}\left( x\right)$ creates) one atom at the position $x$. At
low temperatures atoms experience only small oscillations which can be
described by a field function $u\left( x\right) $ (the function of
displacements). This function can be expanded in modes
\[
u\left( x\right)=\sum_{k}\frac{1}{\sqrt{2\omega \left( k\right) L}}\left(
a_{k}e^{ikx}+a_{k}^{+}e^{-ikx}\right),
\]%
where $L$ is the size of the crystal and coefficients $a_{k}^{+}$ and $a_{k}$
play in quantum theory the role of creation and annihilation operators for
phonons. Thus in the leading order the Hamiltonian takes the form
\begin{equation}
H=\sum_{k}\omega \left( k\right) \left( a_{k}^{+}a_{k}+\frac{1}{2}\right) ,
\label{h}
\end{equation}
where $\omega \left( k\right) $ is the energy of a phonon. The ground state
corresponds to the zero temperature $T=0$ and represents the vacuum state
for phonons, i.e., $a_{k}\left\vert 0\right\rangle =0$, for all $k$.

Consider now the case of a nontrivial topology of the crystal. To this aim
we join two arbitrary points $x_{1}$ and $x_{2}$ of the crystal, separated
by a distance $\ell$, with an additional chain of atoms of a length of the
order $\ell$. The presence of the additional chain of atoms means a
degeneracy that appears in the system: more than one atom may have the same
position $x$. In order to describe such a situation we have two possible
ways. First one is to change the number of dimensions of the system, i.e.,
to introduce an additional coordinate $y$ by means of which different atoms,
which belong to different chains and have the same position $x$, could be
distinguished. This way presumes that the additional coordinate $y$ is an
observable and can be somehow measured. In the example under consideration
atoms are 3-dimensional particles and, therefore, $y$ is simply one of the
coordinates normal to the line of the crystal. For other degenerate systems,
however, $y$ has no such simple interpretation. It can be related to some
extra dimensions or to some internal group space.

The second way is to replace the single-valued functions $\psi \left(
x\right) $ and $u\left( x\right) $ in the region $x_{1}\leq x\leq $ $x_{2}$
with two-valued functions $\psi ^{\alpha }\left( x\right) $ and $u^{\left(
\alpha \right) }\left( x\right) $, $\alpha =1,2$. This can be interpreted as
an introduction of the set of two identical fields. The second approach is
more general than the first one, because it works also in the case when the
extra coordinate $y$ cannot be measured and, therefore, quantum states,
which differ in the extra coordinate only, are physically equivalent and
should be considered as the same state. In particular, it is exactly the
situation which is realized if topology changes in the system are forbidden,
for in this case interaction terms in the Hamiltonian do not contain matrix
elements which may change the extra coordinate (although this does not mean
that the extra coordinate cannot be measured in principle, the removal of
the degeneracy requires to invoke additional matter fields which are not a
priori contained in the system).

In this manner, we can introduce an operator of the number of fields $%
N\left(x\right)$, which in our example represents the characteristic
function $N\left(x\right) =2$ at $x_{1}\leq x\leq x_{2}$ and $N\left(
x\right) =1$ at the rest of the points of the crystal. It is important that
the number of fields operator depends on quantum states (in our example it
is the dependence on $x$). In a more complicated case $N\left( x\right)$ is
an arbitrary integer-valued function which characterizes the degree of
degeneracy at different values of $x$ or, in other words, the structure of
the crystal. The exact form of the structural function depends on conditions
which the system was prepared in. Note that such complex systems have a
rather wide spread in the nature, e.g. fractal media \cite{Fr,Kor} can be
viewed as low-dimensional systems with rather complex structural function $%
N\left( x\right)$. We also point out that in the case of a variable topology
(e.g. in the case of percolation systems \cite{Fr2}) the structural function
$N\left(x\right)$ represents an additional dynamical variable which depends
on time.

In the example above the two fields $u^{1}\left(x\right)$ and $%
u^{2}\left(x\right)$ are identical, for they describe identical atoms. From
the mathematical standpoint this means that the Hamiltonian always inherits
(from the atoms) the symmetry with respect to permutation $u^{\alpha }\left(
x\right) \leftrightarrow u^{\beta}\left(x\right)$. Upon quantizing, fields
are described by sets of particles (or quasi-particles). Whether the
particles of different sorts (i.e. those which differ in the extra
coordinate $y$) are identical or not, relays heavily on the possibility to
measure the extra coordinate. When such measurements are intrinsically
forbidden, quantum states, which differ in $y$ only, are physically
equivalent, hence respective particles should be considered as
indistinguishable. This gives rise to the fact that excitations in such a
crystal obey a generalized statistics.

In Ref. \cite{K99} an analogous picture was proposed in order to account for
possible space-time foam effects. To describe the topology changes during
the very early period of the evolution of the Universe one can fancy
processes where different pieces of the space constantly re-glued at random
points. Once these processes stopped, the structure remains frozen, which
means that at the present stage the physical space comes in a number of
copies glued together. An equivalent statement is simply that every physical
field has to be multivalued. As it was just explained in the example above,
once the topology is no longer changing, the fields defined on different
sheets of the space have to be taken as indistinguishable. Therefore, the
conclusions made above for quasi-particles in a crystal of a non-trivial
topology have to hold true for real particles in an empty space as well. In
this sense, MOFT can be viewed as a specialized version of a field theory
obeying a generalized statistics. Main principles of the generalized
statistics and the generalized second quantization scheme are briefly
described in the next section.

\section{Generalized second quantization scheme and generalized statistics}

Consider a system of identical particles with an undefined a priori symmetry
of wave functions. We shall use the Bogoliubov method \cite{B}, in which the
second quantization is applied to the density matrix (for the case of
para-statistics this approach was extended in Ref. \cite{gov}). Let us
define operators $M_{ij}$ of transitions for particles from a quantum state $%
j$ into a quantum state $i$. These operators must obey the Hermitian
conditions, i.e.
\begin{equation}
M_{ij}^{+}=M_{ji},
\end{equation}
and the algebraic relations expressing the indistinguishability principle
for identical particles:
\begin{equation}
\left[ M_{ij},M_{km}\right] =\delta _{jk}M_{im}-\delta _{im}M_{kj}.
\label{NB}
\end{equation}

Consider now systems with a variable number of particles. To this end we
need to introduce a set of creation and annihilation operators for particles
($a_{i}^{+}$ and $a_{i}$) and somehow express the transition operators $%
M_{ij}$ via them. The simplest generalization of Bose and Fermi statistics
was first suggested by H.S. Green \cite{G53} and later by D.V. Volkov \cite%
{Vol} and is called the parastatistics, or the Green-Volkov statistics.

Consider a set of creation and annihilation operators of particles $%
a_{k}^{+} $ and $a_{k}$, while the transition operators are presented in the
form
\begin{equation}
M_{ik}=\frac{1}{2}\left( a_{i}^{+}a_{k}\pm a_{k}a_{i}^{+}\mp n_{ik}\right) ,
\label{Nd}
\end{equation}
where $n_{ik}$ is, in general, an arbitrary Hermitian matrix. The upper sign
stands for the generalized Bose statistics, while the lower sign stands for
the generalized Fermi statistics. The operator $M_{i}=M_{ii}$ is the
operator of the number of particles in the quantum state $i$. Then the
creation and annihilation operators should obey the requirements
\begin{equation}
\left[ M_{i},a_{k}\right] =-\delta _{ik}a_{k},\ \left[ M_{i},a_{k}^{+}\right]
=\delta _{ik}a_{k}^{+}.  \label{NA}
\end{equation}
Relations (\ref{NB}) and (\ref{NA}) remain invariant under unitary
transformations
\begin{equation}
a_{i}^{\prime }=\sum_{k}u_{ik}a_{k},\ a_{i}^{\prime +}=\sum_{k}
u_{ik}^{\ast}a_{k}^{+},
\end{equation}
where $\sum_{m}u_{im}u_{km}^{\ast }=\delta _{ik}$. Thus, applying to (\ref%
{NA}) an infinitesimal transformation $u_{ik}=\delta _{ik}+\varepsilon
\sigma_{ik}+ o(\varepsilon)$, where $\sigma_{ik}^{\ast }=-\sigma_{ki}$, and
retaining the first order terms in $\varepsilon$ we get the basic
commutation relations for the creation and annihilation operators
\begin{equation}
\lbrack M_{kl},a_{m}^{+}]=\delta _{lm}a_{k}^{+},\ \
[M_{lk},a_{m}]=-\delta_{lm}a_{k},  \label{par-rel}
\end{equation}%
which were first suggested by Green \cite{G53}.

Consider now the vacuum state $\left\vert 0\right\rangle$ that is
\begin{equation}
a_{k}\left\vert 0\right\rangle =0
\end{equation}
for all $k$. Then the requirement that for all $i$ and $k$ the transition
operators annihilate the vacuum state
\begin{equation}
M_{ik}\left\vert 0\right\rangle =0
\end{equation}
leads to the condition on one-particle quantum states in the form
\begin{equation}
a_{k}a_{i}^{+}\left\vert 0\right\rangle =n_{ik}\left\vert 0\right\rangle,
\label{1st}
\end{equation}
which means that the basis of one-particle states is, in general, not
orthonormal, but it has norms $\left\langle 0\right\vert
a_{k}a_{i}^{+}\left\vert 0\right\rangle =n_{ik}$. From the physical
standpoint this signals up the presence of a degeneracy of quantum states
(the presence of the extra coordinate in the example of the previous
section).

Since $n_{ik}$ is a Hermitian matrix, it follows that it has the diagonal
form in a certain basis of one-particle wave functions, i.e. $n_{ik}=\delta
_{ik}N_{k}$ (in the previous section we discussed the case where this matrix
has a diagonal form in the coordinate representation). In the simplest case $%
N_{k}=N$ is a constant and therefore the relation $n_{ik}=\delta _{ik}N$
remains invariant in an arbitrary basis. Then the condition that norms of
vectors in the Fock space are positively defined leads to the requirement
that $N$ is an integer number which characterizes the rank of the
statistics, or the degree of degeneracy of quantum states \cite{G53,gov}. In
this simplest case the number $N$ corresponds to the maximal number of
particles which admit an antisymmetric (for parabosons) or symmetric (for
parafermions) state. The case $N=1$ corresponds to the standard Bose and
Fermi statistics. For the case of a constant rank, Green also gave an ansatz
which resolves the relations (\ref{par-rel}) and (\ref{1st}) in terms of the
standard Bose and Fermi creation and annihilation operators
\begin{equation}
a_{p}^{+}=\sum_{\alpha =1}^{N}b_{p}^{(\alpha )+},\ \ \
a_{k}=\sum_{\alpha=1}^{N}b_{k}^{(\alpha )},  \label{G-r}
\end{equation}%
where $b_{p}^{(\alpha )}$ and $b_{p}^{(\beta )+}$ are the standard Bose
(Fermi) operators at $\alpha =\beta $ (i.e. $\left[ b_{p}^{\left(\alpha
\right)} b_{k}^{\left( \alpha \right)+}\right]_{\pm }=\delta _{pk}$), but
they anti-commutate (commutate) as $\alpha \neq \beta $ (i.e. $\left[
b_{p}^{\left( \alpha \right) }b_{k}^{\left( \beta \right) +}\right] _{\mp
}=0 $) for the case of parabose (parafermi) statistics. The presence of an
additional index $\alpha$ in the creation and annihilation operators removes
the above mentioned degeneracy of one-particle quantum states.

In the coordinate representation, creation and annihilation of particles is
described by the secondly quantized field operators
\begin{equation}
\widehat{\psi }\left( x\right) =\sum \psi _{k}\left( x\right) a_{k},~\
\widehat{\psi }^{+}\left( x\right) =\sum \psi _{k}^{\ast }\left( x\right)
a_{p}^{+},  \label{WF}
\end{equation}
where $\left\{ \psi _{k}\left( x\right) \right\} $ is a basis of normalized
one--particle wave functions. The meaning of these operators is that $%
\widehat{\psi }^{+}\left( x\right) $ creates (while $\widehat{\psi }\left(
x\right) $ annihilates) a particle at the point $x$. In this manner, in the
case of parastatistics of a constant rank, field operators $\widehat{\psi }
\left( x,t\right) $ can be presented in the form $\widehat{\psi }\left(
x,t\right) =\sum_{\alpha =1}^{M}\widehat{\psi }^{(\alpha )}\left( x,t\right)
$, i.e. the particles can, in fact, be described by a set of ordinary fields
$\{\widehat{\psi }^{(\alpha )}\left( x,t\right) \}$, while the many-particle
states are classified by the set of occupation numbers $\left\vert
m_{k}^{\left( \alpha \right) }\right\rangle $. In other words the Green
representation (\ref{G-r}) transforms paraparticles into the set of
particles of different sorts, or equivalently into identical particles with
an additional internal coordinate. Again, the indistinguishability of
particles is here the result of the impossibility of measuring the extra
coordinate, which reflects the symmetry of the Hamiltonian with respect to
permutations of particles of different sorts. Thus, the states which have
the same total occupation numbers $m_{k}=\sum m_{k}^{\left( \alpha \right) }$
should be considered as physically equivalent states.

One can easily generalize the Green representation (\ref{G-r}) onto the case
of an arbitrary Hermitian matrix $n_{ik}$. In the basis in which this matrix
takes the diagonal form ($n_{ik}=N_{k}\delta _{ik}$) the Green
representation is given by the same expression (\ref{G-r}) in which,
however, the rank of statistics $N_{k}$ depends on the quantum state (the
index $k$). Thus, in the general case, the rank of statistics represents an
additional variable. Note that in an arbitrary basis the Green
representation does not work and, therefore, we can say that the matrix $%
n_{ik}$ distinguishes a preferred basis of quantum states. In the previous
section we have interpreted the functional dependence of the rank of
statistics on the state (and hence the matrix $n_{ik}$) as a characteristic
of the topological structure of the system. Thus, we can say that the
topological structure defines a preferred basis of one-particle wave
functions for which the classification of quantum states takes the simplest
form.

In conclusion of this section we note that from the pure mathematical point
of view instead of the Green's ansatz (\ref{G-r}) we can use the standard
commutation rules for operators $b_{p}^{(\alpha )}$ and $b_{p}^{(\beta )+}$,
i.e. $\left[ b_{p}^{\left( \alpha \right) }b_{k}^{\left( \beta \right) +}%
\right] _{\pm }=\delta _{pk}\delta _{\alpha \beta }$. For statistics of a
constant rank ($N=const$) this case is trivial, since it corresponds to the
usual Bose and Fermi statistics, i.e. $\left[ a_{i}a_{k}^{+}\right] _{\pm
}=n_{ik}=N\delta _{ik}$. Still, in applications this case should be
considered on an equal footing with the case of parastatistics. Indeed, in
the example considered in the previous section, there exist fermionic
excitations which obey the standard statistics we just described. For such
excitations, according to the Green ansatz, no more than one fermion can
occupy the same position $x$. In the presence of additional links (chains of
atoms) it is surely not the general case: if the crystal has $N$ different
links at a position $x$, then the maximal number of fermionic excitations
will also be $N$ (which is the number of different fermionic fields). This
means that there will always exist fermionic excitations which obey the
parafermi statistics. In real crystals, the degeneracy (which appears due to
the presence of the extra coordinate) can be removed (e.g., if processes
involving topology transformations are not suppressed), particles with
different extra coordinates can become distinguishable and then the presence
of a set of fields (instead of a single field) will be essential. It is
clear that the same concerns the statistics of bosonic excitations (phonons).

\section{Renormalization of interaction constants}

In the present section we show that independently on the choice of the
statistics of particles the fact that particles in the space of a nontrivial
topological structure are described by a set of identical fields results in
an additional renormalization of all interaction constants \cite{KT02}.
Indeed, let $S$ be a background basic space and let us specify an arbitrary
field $\varphi $ on it. We suppose that the action for the field can be
presented in the following form (for the sake of simplicity we consider the
case of linear perturbations only):
\begin{equation}
I=\int\limits_{S}d^{4}x\left( -\frac{1}{2}\varphi \widehat{L}\varphi +\alpha
J\varphi \right) ,  \label{act1}
\end{equation}%
where $\widehat{L}=\widehat{L}\left( \partial \right) $ is a differential
operator (e.g., in the case of a massive scalar field $\widehat{L}\left(
\partial \right) =\partial ^{2}+m^{2}$), $J$ is an external current, which
is produced by a set of point sources ($J=\sum J_{k}\delta \left(
x-x_{k}(s)\right) $, where $x_{k}\left( s\right) $ is a trajectory of a
source), and $\alpha $ is the value of the elementary charge for sources.
Thus, the field $\varphi $ obeys the equation of motion
\begin{equation}
-\widehat{L}\varphi +\alpha J=0.
\end{equation}%
We note that this is valid for perturbations in gauge theories ($\varphi
=\delta A_{\mu }$, where $\alpha $ is the gauge charge) and in gravity ($%
\varphi =l_{pl}\delta g_{\mu \nu }$, where $\alpha =l_{pl}$ is the Planck
length).

In the Modified Field Theory we admit that space have a nontrivial
topological structure which cannot be a direct observable. This means that
the field $\varphi$ is not a usual field any more, but is a generalized
field which upon quantization gives rise to particles obeying a generalized
statistics. It was demonstrated in the previous two sections that the
topological structure can be defined by a structural matrix $n\left(
x,y\right) $ which determines the degree of the degeneracy of one-particle
quantum states. In the basis $\{f_{i}\}$ in which this matrix takes a
diagonal form $n_{ik}=N_{k}\delta _{ik}$ the nontrivial topological
structure is accounted for by the replacement of the field $\varphi_{k}$ ($%
\varphi =\sum_{k}\varphi_{k}f_{k}$) with a set of fields $\varphi_{k}^{a}$, $%
a=0,1,...,N_{k}$.

In particle physics the momentum representation is commonly used (i.e.,
Fourier transforms $f_{k}\sim \frac{1}{\sqrt{V}}\exp \left( -ikx\right) $),
where the states of the field can be classified in terms of free particles.
In order to allow for the existence of free particles in MOFT, we assume
that the matrix $n$ is diagonal exactly in the momentum representation.
Moreover, we will show later on that in thermodynamic equilibrium the number
of fields $N_{k}$ in the momentum space is a quite natural object. Thus, the
total action assumes the structure
\begin{equation}
I=\int dt\sum_{k}\sum_{a=0}^{N_{k}}\left( -\frac{1}{2}\varphi _{k}^{\ast a}%
\widehat{L}_{k}\varphi _{k}^{a}+\alpha J_{k}^{\ast }\varphi _{k}^{a}\right) .
\label{act2}
\end{equation}%
where $\widehat{L}_{k}=\widehat{L}(\partial _{t},-ik)$. Fields $\varphi
_{k}^{a}$ are supposed to obey the identity principle and, therefore, they
equally interact with the external current.

It is easy to see that the main effect of the introduction of the number of
identical fields is the renormalization of the charge (the constant $\alpha $%
). To this end we introduce a new set of fields as follows
\begin{equation}
\varphi _{k}^{a}=\frac{\widetilde{\varphi }_{k}}{\sqrt{N_{k}}}+\delta
\varphi _{k}^{a},\,\;\;\sum_{a}\delta \varphi _{k}^{a}=0  \label{eff}
\end{equation}%
where $\widetilde{\varphi }_{k}$ is the effective ordinary field \cite{K99}
\begin{equation}
\widetilde{\varphi }_{k}=\frac{1}{\sqrt{N_{k}}}\sum_{a=0}^{N_{k}}\varphi
_{k}^{a}.
\end{equation}%
Then the action splits into two parts
\begin{equation}
I=\int dt\left( -\frac{1}{2}\sum_{k,a}\delta \varphi _{k}^{\ast a}\widehat{L}%
_{k}\delta \varphi _{k}^{a}\right) +\int dt\left( -\frac{1}{2}\sum_{k}%
\widetilde{\varphi }_{k}^{\ast }\widehat{L}_{k}\widetilde{\varphi }%
_{k}+\sum_{k}\widetilde{\alpha }_{k}J_{k}^{\ast }\widetilde{\varphi }%
_{k}\right) .  \label{efact1}
\end{equation}%
The first part represents a set of free fields $\delta \varphi ^{a}$ which
are not involved into interactions between particles and, therefore, cannot
be directly observed. The second part represents the standard action for the
effective field $\widetilde{\varphi }$ with a new value for the charge $%
\widetilde{\alpha }_{k}=\sqrt{N_{k}}\alpha $ which now depends on the wave
number $k$, i.e. it becomes scale-dependent.

We recall that $N_{k}$ is an operator and we, strictly speaking, should
consider an average value for the charge
\begin{equation}
\left\langle \widetilde{\alpha }\left( k\right) \right\rangle =\left\langle
\sqrt{N_{k}}\right\rangle \alpha .  \label{charge}
\end{equation}%
The requirements of homogeneity and isotropy of the Universe allow $%
\left\langle N_{k}\right\rangle =N_{k}\left( t\right) $ to be an arbitrary
function of $|k|$. This means that every point source is distributed in
space with the density
\begin{equation}
\rho \left( r\right) =\frac{1}{2\pi ^{2}}\int\limits_{0}^{\infty }\left(
\sqrt{N_{k}}k^{3}\right) \frac{\sin \left( kr\right) }{kr}\frac{dk}{k}.
\label{dis}
\end{equation}%
If we assume (and we do so) that processes with topology transformations
have stopped after the quantum period in the evolution of the Universe, then
the structure of the momentum space conserves indeed and the function $%
\left\langle N_{k}\right\rangle $ depends on time via only the cosmological
shift of scales, i.e., $\left\langle N_{k}\right\rangle =N_{k\left( t\right)
}$, where $k\left( t\right) \sim 1/a\left( t\right) $ and $a\left( t\right) $
is the scale factor. In this manner, function $N_{k}$ represents some new
universal characteristic of the physical space.

\section{Description of particles in MOFT}

Let $\mathit{\psi }$ be an arbitrary field which, upon the expansion in
Fourier modes, is described by a set of creation and annihilation operators $%
\left\{a_{\alpha ,k},a_{\alpha ,k}^{+}\right\} $, where the index $\alpha $
enumerates polarizations and distinguishes between particles and
antiparticles. In what follows, for the sake of simplicity, we ignore the
presence of the additional discrete index $\alpha $. These operators are
supposed to satisfy the relations
\begin{equation}
a_{k}a_{p}^{+}\pm a_{p}^{+}a_{k}=\delta_{kp},  \label{b}
\end{equation}%
where the sign $\pm $ depends on the statistics of particles. In MOFT the
number of fields is a variable and, therefore, the set of operators $%
\left\{a_{k},a_{k}^{+}\right\}$ is replaced with the expanded set $\left\{
a_{k}\left( j\right) ,a_{k}^{+}\left( j\right) \right\}$, where $j\in \left[
1,...,N_{k}\right] $. For a free field, the energy is an additive quantity,
so it can be written as
\begin{equation}
H_{0}=\sum_{k}\sum_{j=1}^{N_{k}}\omega _{k}a_{k}^{+}\left( j\right)
a_{k}\left( j\right) ,  \label{en}
\end{equation}%
where $\omega _{k}=\sqrt{k^{2}+m^{2}}$. When there is an interaction
described by a potential $V$, the total Hamiltonian $H=H_{0}+V$ can be
expanded, due to the symmetry with respect to the permutation of particles
of different sorts $j$, in the set of operators \cite{K99}
\begin{equation}
A_{m_1,m_2}\left( k\right) =\sum_{j=1}^{N_{k}}\left(a_{k}^{+}\left( j\right)
\right)^{m_1}\left(a_{k}\left( j\right)\right)^{m_2}.  \label{a}
\end{equation}

For bosonic particles, a single field mode (corresponding to a fixed wave
number $k$) is a quantum mechanical oscillator with a countable set of
equidistant energy levels; the level number $n$ corresponds to $n$ bosons
with the wave number $k$. A single fermionic oscillator (i.e. a field mode
for fermionic particles) may have only two states, corresponding to $n=0$
and $n=1$. In a complete theory with a variable number of fields, quantum
states are classified by occupation numbers. To this end, we consider the
set of operators $\left\{ C\left(n,k\right) ,C^{+}\left( n,k\right) \right\}$
which annihilate (resp. create) the oscillators (the field modes) with the
wave number equal to $k$ and the number of particles in the mode equal to $n$%
. From the mathematical standpoint such operators can be constructed by
means of the second quantization of wave functions for field amplitudes $%
\Psi \left( a_{k}\right) $. Here it is important that the amplitude $a_{k}$
represents an observable, i.e. it can be measured. There is no problem with
bosonic fields, for amplitudes of bosonic fields are always good
observables. The problem appears, however, in the case of fermionic fields,
for fermionic amplitudes do not correspond to any observable (all fermionic
observables are bilinear combinations in amplitudes). To overcome this
difficulty we consider a set of "test" fermionic variables (a set of
Grassmanian numbers) $\xi _{k}$ which obey the relations
\begin{equation}  \label{grass}
\xi _{k}\xi _{p}+\xi _{p}\xi _{k}=2\delta _{kp}
\end{equation}
and
\begin{equation}
a_{k}\xi_{p}-\xi _{p}a_{k}=0,
\end{equation}
and construct bilinear combinations $\widetilde{a}_{k}=\xi _{k}a_{k}$. New
amplitudes $\widetilde{a}_{k}$ can already be considered as good observables
and we can define operators $\widehat{\Psi }\left( \widetilde{a}_{k}\right) $
and $\widehat{\Psi }^{+}\left( \widetilde{a}_{k}\right) $ in the standard
way. Therefore, in what follows we will understand under fermionic creation
and annihilation operators the new variables $\widetilde{a}_{k}^{+}$ and $%
\widetilde{a}_{k}$, while the standard fermionic operators can be expressed
as $a_{k}=\xi _{k}\widetilde{a}_{k}$. With this remark the both cases (fermi
and bose fields) can be considered in a unified way.

Thus, the creation and annihilation operators for field modes are supposed
to obey the standard relations
\begin{equation}
C\left( n,k\right) C^{+}\left( m,k^{\prime }\right) \pm C^{+}\left(
m,k^{\prime }\right) C\left( n,k\right) =\delta _{nm}\delta _{kk^{\prime
}}\;.  \label{c}
\end{equation}%
and should be used to construct the Fock space in MOFT. The sign $\pm $ in (%
\ref{c}) depends on the symmetry of the wave function under field
permutations, i.e. on the statistics of fields, which is not the same as the
statistics of the particles. Thus, in the case of bosonic fields both signs
are possible. E.g., in the example of a one-dimensional crystal discussed
earlier, phonons are always bose-particles but the statistics of the phonon
fields $u^{\alpha }$ follows the statistics of atoms. Indeed, phonons
represent nothing more than oscillations of atoms with respect to their
equilibrium positions. The wave function of the crystal is either symmetric
or antisymmetric with respect to permutations of atoms and, therefore, it
will be symmetric or antisymmetric with respect to permutations of phonon
modes (i.e. permutations of the oscillators whose excitations correspond to
the birth of the phonons). In the case of a simple topology of the crystal
the question on the symmetry of phonon modes is not essential (all
oscillators differ by wave numbers, i.e. there is only one oscillator for
every wave number $k$), while in the case of nontrivial topology a larger
than one number of oscillators can correspond to the same wave number $k$
and the type of statistics obeyed by them becomes an issue.

In the case of fermionic fields, however, in order to describe nontrivial
topology we should use the Bose statistics for the oscillators. Indeed, let
us return to the example of a one-dimensional crystal. In the crystal with a
nontrivial topological structure the number of additional links and,
therefore, the number of fields required for the description of excitations
is not restricted. Therefore, there should be no restrictions on the number
of fermionic excitations at a position $x$. It is possible only if fermionic
field modes obey the Bose statistics, because every such mode can support
only one particle (the number of states of a single fermionic oscillator is
bounded by two). Of course, from the formal standpoint the case of Fermi
statistics for fermionic modes should not be excluded as well. This case,
however, corresponds to a standard fermionic field, because the total number
of fermionic oscillators will be restricted by $N\leq 2$, and the case $N=2$
can be identified with $N=0$ (in both cases no particles can be created).

In terms of $C$ and $C^{+}$, operators (\ref{a}) can be expressed as follows
\begin{equation}
A_{m_{1},m_{2}}\left( k\right) =\sum\limits_{n}\frac{\sqrt{\left(
n+m_{1}\right) !\left( n+m_{2}\right) !}}{n!}C^{+}\left( n+m_{1},k\right)
C\left( n+m_{2},k\right)  \label{A}
\end{equation}%
where in the case of bosonic fields the sum is taken over the values $%
n=0,1,..$, while in the case of fermionic fields $n=0,1$, $n+m_1=0,1$ and $%
n+m_2=0,1$. Thus, the eigenvalues of the Hamiltonian of a free field take
the form
\begin{equation}
H_{0}=\sum_{k}\omega _{k} A_{1,1}\left( k\right) =\sum_{k,n\geq
1}n\omega_{k}N\left(n,k\right) ,
\end{equation}%
where $N\left(n,k\right)$ is the number of modes with the wave number $k$
and the number of particles $n$ (i.e., $N\left( n,k\right) =C^{+}\left(
n,k\right) C\left( n,k\right) $).

Thus, the field state vector $\Phi $ is a function of the occupation numbers
$\Phi \left( N\left( n,k\right) ,t\right) $, and its evolution is described
by the Shr\"{o}dinger equation
\begin{equation}
i\partial _{t}\Phi =H\Phi .
\end{equation}%
Consider the operator
\begin{equation}
N_{k}=A_{0,0}\left( k\right) =\sum\limits_{n}C^{+}\left( n,k\right) C\left(
n,k\right)  \label{N(k)}
\end{equation}%
which characterizes the total number of modes for a fixed wave number $k$.
In standard processes when the number of fields is conserved (e.g., when
topology transformations are suppressed) $N_{k}$ is a constant of motion $%
\left[N_{k},H\right] =0$ and, therefore, this operator can be considered as
an ordinary fixed function of the wave numbers.

Consider now the particle creation and annihilation operators. Among the
operators $A_{m_1,m_2}\left( k\right) $ are some which change the number of
particles by one
\begin{equation}
b_{m}^{-}\left( k\right) =A_{m,m+1}\left( k\right) ,\;\;b_{m}^{+}\left(
k\right) =A_{m+1,m}\left( k\right) ,
\end{equation}%
and which replace the standard operators of annihilation and creation of
particles, i.e., they satisfy the relations
\begin{equation}
\left[ \widehat{n},b_{m}^{(\pm )}\left( k\right) \right] =\pm b_{m}^{(\pm
)}\left( k\right) ,\;\;\;\left[ H_{0},b_{m}^{(\pm )}\left( k\right) \right]
=\pm \omega _{k}b_{m}^{(\pm )}\left( k\right) ,
\end{equation}%
where
\begin{equation}
\widehat{n}=\sum_{k}\widehat{n}_{k}=\sum_{k,n}nN\left( n,k\right).
\end{equation}%
In the case of fermions there exist only two such operators $%
b_{0}^{+}\left(k\right)=C^+\left(1,k\right)C\left(0,k\right)$ and $%
b_{0}^{-}\left(k\right)=C^+\left(0,k\right)C\left(1,k\right)$. As we
mentioned, proper (para)fermion creation and annihilation operators are
given by $\xi_k b_{0}^{+}\left(k\right)$ and $\xi_k b_{0}^{-}\left(k\right)$
where $\xi_{k}$ is a Grassmanian number (see (\ref{grass})). Then, in the
case $N_{k}=1$ they transform into the standard fermion creation and
annihilation operators.

In the case of bosons the total number of creation/annihilation operators is
determined by the structure of the interaction term $V$. In the simplest
case (e.g., in the electrodynamics) the interaction term is expressed solely
via $b_{0}^{+}\left( k\right) $ and $b_{0}^{-}\left( k\right) $. In this
case, we can introduce creation/annihilation operators for the effective
field
\begin{equation}
a_{k}^{\prime }=\frac{1}{\sqrt{N_{k}}}b_{0}^{-}\left( k\right)
,\;\;a_{k}^{\prime +}=\frac{1}{\sqrt{N_{k}}}b_{0}^{+}\left( k\right) ,
\label{ef}
\end{equation}%
which satisfy the standard commutation relations, i.e., $\left[
a_{k}^{\prime },a_{p}^{\prime +}\right] =\delta _{kp}$. This restores the
standard theory, but new features appear, however. First, as it was shown in
the previous section, the renormalization (\ref{ef}) results in the
renormalization of interaction constants. Secondly, if the fermionic
oscillators obey Bose statistics, then in the region of wave numbers in
which $N_{k}>1$ fermions violate the Pauli principle: up to $N_{k}$ fermions
can be created with the same wave number $k$.

\section{Vacuum state in MOFT}

In this section we describe the structure of the vacuum state for bosons and
fermions. The true vacuum state in MOFT is defined by the relation
\begin{equation}
C\left( n,k\right) \left\vert 0\right\rangle =0.
\end{equation}%
In this true vacuum state all modes are absent: $N_{k}=0$, hence no
particles can be created and all observables related to the field are
absent. Thus, the true vacuum state corresponds to the absence of physical
space and, in reality, cannot be achieved. Assuming that upon the quantum
period of the evolution of the Universe, topology transformations are
suppressed, we should require that the number of fields conserves in every
mode: $N_{k}=const$ . Then we can define the ground state of the field $\psi$
(which is the vacuum for the particles) as the vector $\Phi _{0}$ satisfying
the relations
\begin{equation}
b_{m}\left( k\right) \Phi _{0}=0  \label{0}
\end{equation}%
for all values $k$ and $m=0,1,...$ . However, these relations still do not
define a unique ground state and should be completed by relations which
specify the distribution of modes $N_{k}$.

Consider first the case of bosons, and let the bosonic modes obey Fermi
statistics (i.e., $\left\{ C\left( n,k\right) C\left( m,p\right) \right\}
=\delta _{nm}\delta _{kp}$). The state $\Phi _{0}$ corresponds to the
minimum energy for a fixed mode distribution $N\left( k\right) $. It can be
characterized by additional relations
\begin{equation}
b_{N_{k}+m}^{+}\left( k\right) \Phi _{0}=0\quad (m=0,1,...),
\end{equation}%
or, equivalently, by the occupation numbers
\begin{equation}
N\left( n,k\right) =\theta \left( \mu _{k}-n\omega _{k}\right) =\theta
\left( N_{k}-1-n\right)   \label{d}
\end{equation}%
where $\theta \left( x\right) $ is the Heaviside step function and $\mu _{k}$
is the chemical potential which is related to the number of modes $N_{k}$ as
\begin{equation}
N_{k}=\sum_{n}\theta \left( \mu _{k}-n\omega _{k}\right) =1+\left[ \frac{\mu
_{k}}{\omega _{k}}\right] .  \label{d00}
\end{equation}%
In particular, from (\ref{d}) we find that the ground state for bosonic
fields contains real particles
\begin{equation}
n_{k}^{\ast }=\sum_{n=0}^{\infty }nN\left( n,k\right) =\frac{1}{2}%
N_{k}\left( N_{k}-1\right)   \label{hp}
\end{equation}%
and, therefore, corresponds to a finite energy $E_{0}=\sum \omega
_{k}n_{k}^{\ast }$. These particles, however, are \textquotedblleft
dark\textquotedblright , for they correspond to the ground state.

In the case of fermionic oscillators which obey Fermi statistics, the
occupation numbers in the ground state satisfy (\ref{d}) as well, with the
total number of modes always bounded as $N_{k}\leq 2$. In the same way as
the state $N_{k}=0$, the state with $N_{k}=2$ can also be considered as a
vacuum state (no particles can be created, no annihilated). However, this
vacuum also includes hidden particles $n_{k}^{\ast }=\theta
\left(N_{k}-2\right)$.

In the case of Bose statistics for modes, the vacuum state is given by the
occupation numbers
\begin{equation}
N\left( n,k\right) =N_{k}\delta _{n,0},  \label{f}
\end{equation}
i.e. it is formally constructed from the true vacuum state as follows
\begin{equation}
\Phi _{0} = \prod_{k}\frac{\left( C^{+}\left( 0,k\right) \right) ^{N_{k}}}{%
\sqrt{N_{k}!}}\left\vert 0\right\rangle ,
\end{equation}
In contrast to the case of Fermi statistics for modes, the ground state (\ref%
{f}) contains no particles and corresponds to the zero energy $E_{0}=0$.

In the case of Fermi particles this ground state satisfies the relation
\begin{equation}
\left( b_{0}^{+}\left( k\right)\right)^{N_{k}+1}\Phi _{0}=0.  \label{r}
\end{equation}
Here, the total number of particles $n_{k}$, which can be created at the
given wave number $k$, takes values $n_{k}=0,1,...,N_{k}$, i.e. it cannot
exceed the number of modes. In this case the basis of the Fock space
consists of vectors of the type
\begin{equation}
\left\vert N_{k}-n_{k},n_{k}\right\rangle =\prod_{k}\sqrt{\frac{N_{k}!} {%
\left(N_{k}-n_k\right)!n_{k}!}}\left( b^{+}\left( k\right) \right)
^{n_{k}}\left\vert N_{k},0\right\rangle ,
\end{equation}%
where $n_{k}=0,1,...,N_{k}$.

Now, assigning a specific value for the function $N_{k}$, expressions (\ref%
{d}) and (\ref{f}) define the ground state for respective particles. The
function $N_k$ itself cannot be defined within the corresponding field
theory. We interpret $N_{k}$ as a geometric characteristic of the momentum
space, which has formed during the quantum period in the evolution of the
Universe. Hence, a rigorous derivation of the properties of the function $%
N_{k}$ requires studying processes involving topology changes during that
period. At the moment, we do not have an exact model describing the
formation of $N_{k}$ and, therefore, our consideration will have a
phenomenological character \cite{KT02,K03}.

\section{Origin of the spectral number of fields}

Assume that upon the quantum period of the evolution of the Universe the
matter was thermalized with a very high temperature. Then, as the
temperature dropped during the early stage of the evolution, the topological
structure of the space (and the spectral number of fields) has tempered and
the subsequent evolution resulted only in the cosmological shift of the
physical scales.

There exist at least two possibilities. The first and the simplest
possibility is the case where processes involving topology changes generate
a unique function $N_{k}$ which is the same for all fields (regardless of
their type). However, the mathematical structure of MOFT reserves the more
general possibility when the formation of the spectral distribution of modes
goes in independent ways for different fields. In this case every particular
field $\psi_{a}$ will be characterized by its own function $N_{a}\left(
k\right)$. Which case is realized in the nature can be determined only by
confrontation with observations, and below we consider both cases.

Upon the quantum period, the Universe is supposed to be described by the
homogeneous metric of the form
\begin{equation}
ds^{2}=dt^{2}-a^{2}\left( t\right) dl^{2},
\end{equation}%
where $a\left( t\right) $ is the scale factor, and $dl^{2}$ is the spatial
interval. It is expected that the matter was thermalized with a very high
temperature $T>T_{Pl}$ where $T_{Pl}$ is the Planck temperature. Then the
state of any field was characterized by the thermal density matrix with mean
values for occupation numbers
\begin{equation}
\left\langle N\left( k,n\right) \right\rangle =\left( \exp \left( \frac{%
n\omega _{k}-\mu _{k}}{T}\right) \pm 1\right) ^{-1},  \label{on}
\end{equation}%
where the signs $\pm$ correspond to the choise of statistics (Fermi or Bose)
for the field modes, and the chemical potential $\mu_{k}$ for the given
field is related to the spectral number of field modes as
\begin{equation}
N_{k}=\sum_{n}\left( \exp \left( \frac{n\omega_{k}-\mu _{k}}{T}\right) \pm
1\right) ^{-1}.  \label{N}
\end{equation}

Consider now the first case when the spectral number of fields $N_{k}$ is a
unique function for all fields. It is well known that near the singularity
the evolution of the Universe is governed by a scalar field, while all other
fields can be neglected. We assume that the same field is responsible for
topology transformation processes which took place in the early Universe.
Thus, we can expect that the state of the scalar field was characterized by
the thermal density matrix (\ref{on}) with $\mu =0$ (for the number of
fields varied; strictly speaking, if the field modes satisfy Bose
statistics, the chemical potentials cannot vanish and have to take some rest
value -- one can choose $\mu _{k}=\omega _{k}/2$). On the early stage $m\ll T
$, and the temperature and the energy of scalar particles decreased with
time proportionally to $a(t)^{-1}$. When the temperature droped below a
critical value $T_{\ast }$, which corresponds to the moment $t_{\ast }\sim
t_{pl}$, topological structure (and the number of fields) tempered. This
generated (see (\ref{on})) the value $N_{k}\sim T_{\ast }/\omega _{k}$ for
the case of Fermi statistics of the oscillators of the scalar field, or $%
N_{k}\sim \ln (T_{\ast }/\omega _{k})T_{\ast }/\omega _{k}$ in the Bose
statistics case. The logarithmic factor will not influence the further
results in a noticeable way, so we will further stick to the Fermi
statistics case, for simplicity.

Thus, the value of $N_k$ become frozen as $N_k \sim T_{\ast}/\omega_k$ at $%
t\sim t_{\ast}$, and (\ref{N}) defines the chemical potential for the scalar
field as $\mu \sim T_{\ast}$, a constant for all $k$. Let us neglect the
temperature corrections, which are essential only at $t\sim t_{\ast }$ and
whose role is in smoothing the real distribution $N_{k}$. Then at the moment
$t\sim t_{\ast }$ the ground state of the scalar field will be described by (%
\ref{d}) with $\mu _{k}=\mu =const\sim T_{\ast }$. During the subsequent
evolution, the physical scales are subjected to the cosmological shift,
however the form of this distribution in the comoving frame must remain the
same. Thus, on the later stages $t\geq t_{\ast }$, we find (see (\ref{d00}%
)):
\begin{equation}
N_{k}=1+\left[ \frac{\widetilde{k}_{1}}{\Omega _{k}\left( t\right) }\right]
,\,\,  \label{NN}
\end{equation}%
where $\Omega _{k}\left( t\right) =\sqrt{a^{2}\left( t\right) k^{2}+%
\widetilde{k}_{2}^{2}}$, $\widetilde{k}_{1}\sim a_{0}\mu $, and $\widetilde{k%
}_{2}\sim a_{0}m$ ($a_{0}=a\left( t_{\ast }\right) $). From (\ref{NN}) we
see that there is a finite interval of wave numbers $k\in \lbrack k_{\min
}\left( t\right) ,k_{\max }\left( t\right) ]$ on which the number of fields $%
N_{k}$ changes its value from $N_{k}=1$ (at the point $k_{\max }$) to the
maximal value $N_{\max }=1+\left[ \widetilde{k}_{1}/\widetilde{k}_{2}\right]
$ (at the point $k_{\min }$). The boundary points of the interval of $k$
depend on time and are expressed via the free phenomenological parameters $%
\widetilde{k}_{1}$ and $\widetilde{k}_{2}$ as follows
\begin{equation}  \label{kmm}
k_{\max }=\frac{1}{a\left( t\right) }\sqrt{\widetilde{k}_{1}^{2}-\widetilde{k%
}_{2}^{2}},\;\;k_{\min }=\frac{1}{a\left( t\right) }\sqrt{\widetilde{k}%
_{1}^{2}/\left( N_{\max }-1\right) ^{2}-\widetilde{k}_{2}^{2}}.
\end{equation}
Out of this interval, the number of fields remains constant i.e., $%
N_{k}=N_{\max }$ for the range $k\leq k_{\min }\left( t\right) $ and $%
N_{k}=1 $ for the range $k\geq k_{\max }\left( t\right) $. From restrictions
on parameters of inflationary scenarios we get $m\lesssim 10^{-5}m_{Pl}$ \
which gives $N_{\max }\gtrsim 10^{5}T_{\ast }/m_{pl}$, where $T_{\ast }$ is
the critical temperature at which topology has been tempered. Since we
assume that the numbers $N_k$ are the same for all fields, we can now
substitute (\ref{NN}) in (\ref{N}) and find the corresponding values of the
chemical potentials $\mu _{k}$ for all other particles.

Consider now the second case when the spectral number of modes $N_{k}$ forms
independently for different fields. We choose Fermi statistics for field
oscillators. In this case bosonic fields are described by the same
distribution (\ref{NN}) in which, however, the parameters $\widetilde{k}_{1}$
and $\widetilde{k}_{2}$ are free phenomenological parameters which are
specific for every particular field. Thus, for massless fields we find $%
\widetilde{k}_{2}=k_{\min }=0$ and $N_{k}=1+\left[ k_{\max }/k\right] $. In
the case of fermions the chemical potential $\mu _{k}$ cannot vanish and for
$T\geq T_{\ast }$ it should take some rest value $\mu _{k}=\epsilon_{0}$.
Thus, in the same way as in the case of bosons, we find $N_{k}=1+\theta
\left( k_{\max }-k\right)$, where $\theta \left( x\right) $ is the Heaviside
step function and $k_{\max }\sim T_{\ast }a\left( t_{\ast }\right) /a\left(
t\right) $. In this case the spectral number of fermions is characterized by
the only phenomenological parameter and $N_{k}=N_{\max }=2$ as $k<k_{\max }$.

As we see, the properties of the spectral number of fields can be different,
depending on which case is realized in the nature. We note, however, that if
in the first case the spectral number of fields $N_{k}$ can be considered as
a new geometric characteristics which straightforwardly defines properties
of the space and hence of all matter fields, in the second case we, strictly
speaking, cannot use such an interpretation. Moreover, if the last case is
really realized in the nature, it should relate to yet unknown processes.
Therefore, in the next sections we will discuss the first possibility only.

In conclusion of this section we note that the real distribution can be
different from (\ref{NN}), which depends on the specific picture of topology
transformations in the early Universe and requires the construction of the
exact theory (in particular, thermal corrections smoothen the step-like
distribution (\ref{NN})). However we believe that the general features of $%
N_{k}$ will remain the same.

\section{The law of gravity}

The dependence of charge values upon wave numbers means that particles lose
their point-like character and this leads to the fact that the standard
expressions for Newton's and Coulomb's energy of interaction between
particles break down. In this section we consider corrections to Newton's
law of gravity (corrections to Coulomb's law are identical). Here, the
interaction constant $\alpha \sim m\sqrt{G}$ (where $m$ is the mass of a
particle), and MOFT gives $G\rightarrow G\left( k\right) =N_{k}G$. To make
estimates, we note that at the moment $t\sim t_{\ast }$ the mass of scalar
particles should be small as compared with the chemical potential (which has
the order of the Planck energy), which gives $\widetilde{k}_{1}\gg
\widetilde{k}_{2}$ in (\ref{NN}). Then, in the range $k_{\max }\left(
t\right) \geq k$ $\gg k_{\min }\left( t\right) $ the function $N_{k}$ can be
approximated by
\begin{equation}
N_{k}\sim 1+\left[ k_{\max }\left( t\right) /k\right] .  \label{mv}
\end{equation}

Consider two rest point particles with masses $m_{1}$ and $m_{2}$. Then the
Fourier transform for the energy of the gravitational interaction between
particles is given by the expression
\begin{equation}
V\left( \mathbf{k}\right) =-\frac{4\pi Gm_{1}m_{2}}{\left\vert \mathbf{k}%
\right\vert ^{2}}N_{k}.  \label{80}
\end{equation}%
The coordinate representation is given by the integral
\begin{equation}
V\left( r\right) =\frac{1}{2\pi ^{2}}\int\limits_{0}^{\infty } V\left(
\omega \right) \omega ^{3} \frac{\sin \left( \omega r\right) }{\omega r}%
\frac{d\omega }{\omega }.  \label{90}
\end{equation}%
From (\ref{NN}) and (\ref{kmm}) we find that this integral can be presented
in the form
\begin{equation}
V\left( r\right) =-\frac{2Gm_{1}m_{2}}{\pi }\sum\limits_{n=0}^{N_{\max
}-1}\int\limits_{0}^{k_{n}}\frac{\sin \left( \omega r\right) }{\omega r}%
d\omega =-\frac{Gm_{1}m_{2}}{r}\left( 1+\sum\limits_{n=1}^{N_{\max }-1}\frac{%
2{Si}\left( k_{n}r\right) }{\pi }\right)  \label{NW}
\end{equation}%
where $k_{n}=\frac{1}{a\left( t\right) n}\sqrt{\widetilde{k}_{1}^{2}-n^{2}%
\widetilde{k}_{2}^{2}}.$ The first term (with $n=0$) of the sum in (\ref{NW}%
) gives the standard expression for Newton's law of gravity, while the terms
with $n>1$ describe corrections. In the range $k_{1}r=k_{\max }r\ll 1$ we
have $Si\left( k_{n}r\right) \sim k_{n}r$, and corrections to Newton's
potential give a constant
\begin{equation}
\delta V\sim -\frac{2Gm_{1}m_{2}}{\pi }\sum\limits_{n=1}^{N_{\max }-1}k_{n}.
\end{equation}%
Thus, in this range we have the standard Newton's force. In the range $%
k_{\min }r\gg 1$, we get $\frac{2}{\pi }Si\left( k_{n}r\right) \sim 1$, and
for the energy (\ref{NW}) we find
\[
V\left( r\right) \sim -\frac{G^{\prime }m_{1}m_{2}}{r},
\]%
where $G^{\prime }=GN_{\max }$. Thus, on scales $r\gg 1/k_{\min }$ the
Newton's law is restored, however the gravitational constant increases in $%
N_{\max }$ times. In the intermediate range $1/k_{\min }\gg r\gg 1/k_{\max }$
the corrections can be approximated as
\begin{equation}
\delta V\left( r\right) \sim \frac{2Gm_{1}m_{2}}{\pi }\frac{\widetilde{k}_{1}%
}{a\left( t\right) }\ln \left( \frac{\widetilde{k}_{2}}{a\left( t\right) }%
r\right) ,  \label{110}
\end{equation}%
i.e., they have a logarithmic behavior.

We note, that from the dynamical point of view the modification of the
Newton's law of gravity can be interpreted as if sources acquire an
additional distribution in space. Indeed, let $m_{1}$ be a test particle
which moves in the gravitational field created by a source $m_{2}$. Then,
assuming Newton's law is unchanged and the test particle is point-like, from
(\ref{NW}) we conclude that the source $m_{2}$ is distributed in space with
the dynamical density
\begin{equation}
\rho _{dyn}\left( r\right) =\frac{m_{2}}{2\pi ^{2}}\int\limits_{0}^{\infty }
N_{k}k^{3} \frac{\sin \left( kr\right) }{kr}\frac{dk}{k}=m_{2}\left( \delta
\left( \vec{r}\right) +\frac{1}{2\pi ^{2}}\sum\limits_{n=1}^{N_{\max }-1}%
\frac{\sin \left( k_{n}r\right) -k_{n}r\cos \left( k_{n}r\right) }{r^{3}}%
\right) .  \label{dark0}
\end{equation}%
Then the total dynamical mass contained within a radius $r$ is
\begin{equation}
M_{dyn}\left( r\right) =4\pi \int_{0}^{r}s^{2}\rho \left( s\right)
ds=m_{2}\left( 1+\frac{2}{\pi }\sum\limits_{n=1}^{N_{\max }-1}\left(
Si\left( k_{n}r\right) -\sin \left( k_{n}r\right) \right) \right) .
\label{m}
\end{equation}%
We stress that this dynamical mass accounts for both distributions (i.e.,
the actual distribution of the source and that of the test particle). Thus,
in the range $r\ll 1/k_{\max }$ we find $M_{dyn}\left( r\right) \sim m_{2}$,
i.e., one may conclude that the gravitational field is created by a point
source with the mass $m_{2}$. However in the range $1/k_{\min }>r>1/k_{\max
} $ the dynamical mass increases as $M_{dyn}\left( r\right) \sim
m_{2}k_{\max }r $, and for $r\gg 1/$ $k_{\min }$ the mass reaches the value $%
M_{dyn}\left( r\right) \sim m_{2}k_{\max }/k_{\min }$. This is in a very
good agreement with what rotation curves in galaxies show \cite{CF}. We note
that the same formulas (\ref{dark0}) and (\ref{m}) work in the case of
charged particles as well, with the obvious substitution $m\sqrt{G}%
\rightarrow e$, which gives rise to the concept of the dynamical charge.

In this manner we see that in MOFT the distributions of the dark matter and
the actual matter are strongly correlated (by the rule (\ref{dark0})), and
the resulting behavior of the dynamically determined mass $M(r)$ agrees with
the observation on the scale of galaxies. We stress that the theoretical
scheme of MOFT was not invented to fit the dark matter distribution. On the
contrary, the logarithmic behavior of the effective field potentials simply
appears in the thermodynamically equilibrium state at the low temperature,
as a by-product of a non-trivial structure of MOFT vacuum.

\section{Dark matter or fictitious baryons}

We see that in the case when the number of fields is conserved MOFT reduces
to the standard field theory in which interaction constants undergo a
renormalization and, in general, acquire a dependence on spatial scales.
From the physical standpoint such a renormalization means that particles
lose their point-like character and acquire a specific distribution in
space, i.e. each point particle is surrounded with an additional halo. This
halo carries charges of all sorts and its distribution around a point source
follows properties of the vacuum state in MOFT (i.e. of the topological
structure of space).

For applied problems the presence of additional halos can be accounted for
by the formal replacement of interaction constants $\alpha $ with operators $%
\widehat{\alpha }\left( -i\nabla \right) $ (where $\nabla =\partial
/\partial x$). In the reference system at rest, which is distinguished by
the microwave background, the Fourier transforms of these operators are
given by the expression $\widetilde{\alpha }\left( k\right) =\sqrt{N_{k}}%
\alpha $ with a universal structural function $N_{k}$. In the simplest case
this function can be taken from (\ref{NN}) and it can be described by two
phenomenological parameters which represent the two characteristic scales.
They are the minimal scale $r_{\min }=2\pi /k_{\max }$ on which the dark
matter starts to show up (and on which the law of gravity (\ref{NW}) starts
to deviate from Newton's law) and the maximal scale $r_{\max }=2\pi /k_{\min
}$ which defines the fraction of the dark matter or the total increase of
interaction constants $G_{\max }=Gr_{\max }/r_{\min }$ (and after which the
Newton's law restores). Both scales depend on time via the cosmological
shift of scales $r\left( t\right) =a\left( t\right) r$. The minimal scale $%
r_{\min }$ can be easily estimated (e.g., see Ref.\cite{kin}) and
constitutes a few $kpc$. To get analogous estimate for the maximal scale $%
r_{\max }$ is not so easy. This requires the exact knowledge of the total
matter density $\Omega _{tot}$ for the homogeneous background and the
knowledge of the baryon fraction $\Omega _{b}$ which gives $r_{\max
}/r_{\min }\sim \Omega _{tot}/\Omega _{b}$ (where $\Omega =\rho /\rho _{cr}$
and $\rho _{cr}$ is the critical density).

If we accept the value $\Omega _{tot}\sim 1$ (which is predicted by
inflationary scenarios) and take the upper value for baryons $\Omega
_{b}\lesssim 0.03$ (which comes from the primordial nuclearsynthesis), we
find $r_{\max }/r_{\min }\gtrsim 30$. Another estimate can be found from
restrictions on parameters of inflationary scenarios. Indeed, in
inflationary models correct values for density perturbations give the upper
boundary for the mass of the scalar field $m\lesssim 10^{-5}m_{Pl}$ \ which
gives $r_{\max }/r_{\min }\gtrsim 10^{5}T_{\ast }/m_{pl}$, where $T_{\ast }$
is the critical temperature at which topology has been tempered. All these
estimates are model dependent, however.

We also note that the more complex ground state can correspond to the
situation in which the topological structure is not homogeneous (which
requires studying the topology formation processes in the early Universe).
In this case the minimal scale can possess some variation in space. Such
variations however should have a characteristic wavelengths $\geq $ $\ell
_{0}$ (at the moment when topology has been tempered $\ell _{0}$ corresponds
to the horizon size), while the structural function should be described in a
mixed representation $N_{k}\left( x\right) $, with $k>2\pi /\ell _{0}$. In
other words, the exact behavior of the structural function should be derived
from observations.

The violation of the gravity law (the logarithmic behavior in a range of
scales) has rather important consequences for the structure formation of the
Universe. It turns out that in the range $r_{\min }\lesssim r\lesssim $ $%
r_{\max }$ (in the case $r_{\max }/r_{\min }\gg 1$) the homogeneous
distribution of matter is unstable \cite{K02}. And this instability works
from the very beginning of the evolution of the Universe. To illustrate the
existence of such an instability we consider the case when the maximal scale
is absent $r_{\max }\rightarrow \infty $, or at least $r_{\max }\gg R_{H}$,
where $R_{H}$ is the Hubble radius. The number of baryons contained within a
ball of a radius $r$ in the case of a uniform distribution with a density $%
n_{b}$ is given by $N_{b}\left( r\right) \sim n_{b}r^{3}$, while the
dynamical mass or charge of every baryon increase according to (\ref{m}) as $%
q_{b}\left( r\right) \sim q_{b}r/r_{\min }$ ($q_{b}$ is the proton mass $%
m_{p}$ or the electric charge $e$). For the total dynamical mass and charge
contained within the radius $r$ we get
\begin{equation}
Q_{dyn}\left( r\right) =q_{b}\left( r\right) N_{b}\left( r\right) +\delta
Q\left( r\right)   \label{ch}
\end{equation}%
where $\delta Q\left( r\right) $ accounts for the contribution of baryons
from the outer region\footnote{%
In the case of homogeneous distribution of baryons this contribution gives $%
Q_{dyn}\left( r\right) \simeq r_{\max }/r_{\min }$ $q_{b}N_{b}\left(
r\right) $ and the total dynamical mass is divergent in the limit $r_{\max
}\rightarrow \infty $.} which does exist according to (\ref{dark0}). Thus,
for the dynamical mass we find $M_{dyn}\left( r\right) \geq \rho
_{b}r^{4}/r_{\min }$, where $\rho _{b}=m_{p}n_{b}$. This means that the
lower limit for the total dynamical density increases with the radius $\rho
_{dyn}$ $\sim $ $M_{dyn}/r^{3}$ $\geq $ $\rho _{b}r/r_{\min }$ and for
sufficiently large $r$ $\sim r_{cr}$ it will reach the value $\rho
_{b}r_{cr}/r_{\min }=$ $\rho _{cr}$ , i.e., $\Omega _{tot}>1$ and the ball
of the radius $r_{cr}$ starts to collapse (i.e., such a Universe must
correspond to a closed cosmological model).

Consider now a small perturbation in the distribution of baryons. From (\ref%
{ch}) we find that the contribution of the perturbation in the total density
will acquire the factor $\geq L/r_{\min }$, where $L$ is the characteristic
size of the perturbed region. For perturbations with $L/r_{\min }\gg 1$ this
increase is considerable\footnote{%
In fact the total increase will be $\sim r_{\max }/r_{\min }$.} and this
will cause the decay of any initial homogeneous distribution which will last
until baryons form a new stable equilibrium distribution.

It turns out that the stable equilibrium distribution of baryons is reached
by a fractal law $N\left( r\right) \sim r^{D}$ with $D\approx 2$. Indeed,
let us suppose that every baryon remains a point-like particle, while the
contribution of its halo to the total dynamical mass and charge can be
accounted for by the presence of some additional number of particles
(fictitious particles). Then we can consider the "dynamical" (or effective)
number of particles $\widetilde{N}_{b}=Q_{dyn}\left( r\right) /q_{b}$ and
the stable equilibrium state will correspond to the homogeneous distribution
of the dynamical charge and mass densities, i.e., $\widetilde{n}_{b}\sim
\widetilde{N}_{b}/r^{3}=const$, while the actual number of baryons in the
range $r_{\min }\lesssim r\lesssim $ $r_{\max }$, as it can be seen from (%
\ref{ch}), will follow the law $N_{b}\left( r\right) \leq Q_{dyn}\left(
r\right) /q\left( r\right) $ $\sim $ $n_{b}r^{2}r_{\min }$. In particular,
this equilibrium state is consistent with the observed homogeneity of the
Universe and the absence of large $\Delta T/T$ (the presence of fictitious
baryons is essential here, for they give at the moment of recombination $%
T^{3}$ $\sim $ $\widetilde{n}$ $=$ $const$), and it perfectly fits the
observed galaxy distribution (see Ref. \cite{LMP}). Above the scale $r_{\max
}$ the\ law $1/r$ restores and the actual distribution of baryons has to
cross-over to homogeneity $N_{b}\left( r\right) \sim r^{3}$. However the
dynamical number of baryons will be $\widetilde{N}_{b}\sim N_{b}r_{\max
}/r_{\min }$. The results of Ref. \cite{LMP} provide that the size of the
upper cutoff $r_{\max }$, if it really exist, must be more than $200Mpc$
and, therefore, for the actual baryon fraction we get a new estimate $\Omega
_{tot}/\Omega _{b}\sim r_{\max }/r_{\min }\geq 10^{5}$. We also recall that,
in fact, due to the large amount of fictitious baryons which have the
homogeneous distribution, the observational limits on $\Delta T/T$ does not
set any restrictions on possible values of $r_{\max }$ and it can be larger
than the Hubble radius.

The instability of a homogeneous distribution of particles means that in
thermodynamic equilibrium baryons were distributed with the fractal law (in
particular, this is the only possibility when $r_{\max }\rightarrow \infty $%
), which gives a homogeneous distribution for the total dynamical mass and
charge (baryons plus fictitious particles) densities. And the observed large
scale structure of the Universe is nothing more than a remnant of the
primordial thermodynamic equilibrium state (which is still in equilibrium).
Upon the recombination, the homogeneous background charge disappears and the
remnant of this process is seen now as the isotropic CMB radiation.

There exists one more argument in favour of this picture. It is the presence
of a diffuse component of the X-ray background \cite{x-r}. Indeed, some
fraction of baryons is, at present, in the hot X-ray emitting intracluster
gas that traces the primordial fractal distribution of baryons. Therefore,
the total dynamical charge (baryons plus fictitious particles) forms a
homogeneous background (i.e., homogeneous hot plasma) in space. In this
sense the X-ray background has very similar origin as that of the cosmic
microwave background radiation.

In this manner we see that dark matter, which in the above picture is
represented by fictitious particles, is actually not dark, for it
contributes to the X-ray background and CMB. Therefore, we should ask
whether fictitious particles can be considered as real particles. The
fictitious character of such particles displays itself in the fact that they
show up only at scales $\ell >r_{\min }$ and, besides, they do not introduce
additional degrees of freedom. E.g., density perturbations of fictitious
particles are caused only by perturbations in the actual baryon density. We
note that at scales $\ell >r_{\min }$ an arbitrary fluctuation in the baryon
density is enhanced by fictitious particles up to the factor $\sim r_{\max
}/r_{\min }$ and, therefore, observed $\Delta T/T$ can be caused by primeval
thermal fluctuations. This also means that at scales $\ell >r_{\min }$ the
standard dust-like equation of state for baryons is invalid (for baryons are
actually not free but strongly interact) and we should expect the appearence
of new phenomena, e.g., this may be used to explain the so-called dark
energy (or quintessence) problem.There is also the possibility to attract
fictitious particles to explain origin of galactic magnetic fields by the
standard mechanism \cite{Har}. However these problems require a deeper
investigation.

The renormalization of all interaction constants means that analogous
background dynamical densities should exist for charges of all sorts. The
concept of a homogeneous background gauge charge is not new, but it has been
already suggested in particle physics by Higgs \cite{Higgs}. In the next
section we discuss how such densities form the rest mass spectrum of
elementary particles.

\section{Origin of the rest mass spectrum of elementary particles}

As it was pointed out above, the renormalization of interaction constants
leaves a rather important imprint in particle physics. We show here that
MOFT allows to relate observed rest mass spectrum of elementary particles to
cosmological parameters (background charge densities), and this can be used
to get an independent estimate for the maximal scale $r_{\max }$. All
particles are supposed to be massless on the very fundamental level. In the
vacuum state the nontrivial topological structure of space produces some
very small values for rest masses of particles (of the order $m_{0}\sim
k_{\max }=2\pi /r_{\min }$, e.g., see Ref. \cite{K00}) and renormalizes the
naked charge values $g\rightarrow \widehat{g}$ (for Fourier transforms we
get $\widetilde{g}\left( k\right) =\sqrt{N_{k}}g$). The present Universe,
however, is not in the vacuum state but can be characterized by a stable
equilibrium state which is described by some background distribution of
charges. Then the interaction of elementary particles with such a background
forms the observed rest mass spectrum. We note that in the standard field
theory such masses are still very small and this is the reason of why we
need to attract additional scalar Higgs fields \cite{Higgs} (Higgs fields
carry gauge charges and, therefore, a constant field produces a homogeneous
background of a gauge charge density). In MOFT, however, interaction
constants increase considerably with scales (e.g., in the region $r_{\min
}\leq r\leq r_{\max }$ the dynamical charge behaves as $q\left( r\right)
\sim q~r/r_{\min }$) and, therefore, masses obtained can be close to
observed values.

Indeed, in the case of fermions the field equation is $\left( \gamma
^{i}\left( \partial _{i}+gA_{i}\right) +m_{F}\right) \psi =0$, where $\gamma
^{i}$ is the Dirac matix, $m_{F}$ is the naked value for the mass of a
fermion, and $A_{i}$ and $\psi $ should be understood as generalized fields.
Then the interaction with a background gauge field $A$ forms the rest mass
value $\delta m_{F}^{2}$ $=$ $\left\langle \widehat{g}A\widehat{g}%
A\right\rangle $ $\sim $ $g^{2}\widetilde{n}_{A}/m_{A}^{0}$, where $g$ is
the naked value of the gauge charge, $m_{A}^{0}\sim k_{\max }$ is the vacuum
(or naked) value of the rest mass for gauge bosons, and $\widetilde{n}$ is
the effective "dynamical" density of the background bosons which accounts
for the renormalization of the charge (by the factor $\sqrt{N_{k}}$) (or the
density of fictitious particles). It is important that the real density of
the bosons does not coincides with $\widetilde{n}_{A}$ and is much smaller
than it. At scales $r\gg r_{\max }$ (i.e., $k\ll k_{\min }$) the fractal
picture crosses over to homogeneity and we can consider bosons to have a
homogeneous distribution with a density $n_{A}$, while the total increase of
the charge is $\widehat{g}^{2}$ $\sim g^{2}$ $r_{\max }/r_{\min }$ and,
therefore, we get for the rest mass the estimate $m_{F}^{2}\sim
g^{2}n_{A}~\left( r_{\max }/r_{\min }\right) /m_{A}^{0}$. Thus, we see that
the observed tiny value of the gauge boson density (e.g., in the case of the
electromagnetic field $n_{\gamma }\sim T_{\gamma }^{3}$, where $T_{\gamma }$
is the CMB temperature) increases by the large factor $r_{\max }/r_{\min }$.
We point out that the actual rest mass values depend on the densities of
fictitious particles $\widetilde{n}_{A}$ which are not direct observables
(on the contrary to $n_{A}$). In particular, in the case when the maximal
scale is absent ($r_{\max }\rightarrow \infty $) the actual particles are
distributed with the fractal law $N_{A}\left( r\right) \sim r^{D}$ which
works for arbitrary large distances and the mean density vanishes $%
n_{A}\rightarrow 0$. However, we still can assign specific finite values to
the densities of fictitious particles $\widetilde{n}_{A}$.

The actual value for the rest mass of a fermion depends on the number of
different interactions (gauge fields) it is involved in, for all
interactions produce independent contributions to the rest mass. E.g.,
neutrinos are the lightest particles, for they are involved in the weak
interaction only, the electron rest mass forms mostly due to the
electromagnetic interactions, while the leading contribution to the rest
mass of baryons (protons and neutrons) comes from the strong interactions.
And indeed, the observed ratios of respective masses give roughly ratios of
gauge charges.

In the case of gauge bosons there exist two independent contributions to the
rest mass. The first comes from the self-interaction (for non-abelian
fields), i.e., from the interaction of bosons with its own background and
this contribution coincides with that for fermions. Thus, if such a
contribution were alone the rest mass spectrum in the case of bosons would
not be different from that of fermions. However, bosons interact as well
with the background formed by fermions (e.g., in the case of photons only
this kind of contribution does exist) and this deviates the resulting
spectrum. E.g., in the case of photons the existence of the background hot
X-ray emitting plasma gives rise to a non-vanishing photon rest mass $%
m_{\gamma }^{2}\sim 4\pi e^{2}\widetilde{n}_{e}/m_{e}$ (which is the
Langmuir frequency), where $\widetilde{n}_{e}=\widetilde{n}_{p}$ is the
background density of fictitious electrons.

We see that in MOFT the background densities of fictitious particles play
the role of Higgs fields. In this picture Higgs fields represent
phenomenological fields which account for the presence of the background
gauge charge and this explains why we do not see Higgs particles in
laboratory experiments. The fictitious particle number densities $\widetilde{%
n}_{i}$ include the renormalization of charge values (the contribution from
halos) and, therefore, are not direct observables (by the astrophysical
means of course; we do observe the rest mass spectrum). This means that the
theory includes the same number of parameters to get the correct rest mass
spectrum as the standard model in particle physics does (we mean here the
number of essential parameters, for such parameters as masses of Higgs
bosons are basically free parameters and are not fixed in the standard
model). However in MOFT such parameters acquire a new astrophysical status.

Now, to make evaluations clear we consider as an example a real scalar field
$\varphi $. Consider the expansion of the field $\varphi $ in plane waves,
\begin{equation}
\varphi \left( x\right) =\sum_{k}\left( 2\omega _{k}L^{3}\right)
^{-1/2}\left( a_{k}e^{ikx}+a_{k}^{+}e{\ }^{-ikx}\right) ,  \label{field}
\end{equation}%
where $\omega _{k}=\sqrt{k^{2}+m_{0}^{2}}$ and $m_{0}$ is a naked value for
the rest mass of scalar particles. In MOFT the field $\varphi \left(
x\right) $ represents a generalized field which means that the creation and
annihilation operators $a_{k}^{+}$ and $a_{k}$ obey the relations (\ref{NA}%
). In particular, for the vacuum state we get (e.g., see (\ref{1st}))
\[
\left\langle 0,N_{k}\right\vert a_{k}a_{i}^{+}\left\vert
N_{k},0\right\rangle =n_{ik}=N_{k}\delta _{ki},
\]%
hence we find for fluctuations of the field potentials in the vacuum state
\begin{equation}
\left\langle \varphi \left( x\right) \varphi \left( x+r\right) \right\rangle
=\frac{1}{\left( 2\pi \right) ^{2}}\int\limits_{0}^{\infty }\frac{dk}{\omega
_{k}}\frac{\sin kr}{kr}\Phi ^{2}\left( k\right) ,  \label{poten}
\end{equation}%
where $\Phi ^{2}\left( k\right) =k^{2}N_{k}$.

Let the potential of the scalar field be $V=\frac{\lambda }{4!}\varphi ^{4}$%
. Then the observed value of the rest mass of scalar particles will be $%
m_{ph}^{2}=m_{0}^{2}+\frac{\lambda }{2}\left\langle \varphi ^{2}\left(
x\right) \right\rangle _{reg}$, where $m_{0}^{2}$ is the initial mass value
(which corresponds to the trivial topology case $N_{k}=1$) and $\left\langle
\varphi ^{2}\left( x\right) \right\rangle _{reg}$ is the regularized mean
value $\left\langle \varphi ^{2}\left( x\right) \right\rangle
_{reg}=\left\langle 0,N_{k}\right\vert \varphi ^{2}\left( x\right)
\left\vert N_{k},0\right\rangle -\left\langle 0,1\right\vert \varphi
^{2}\left( x\right) \left\vert 1,0\right\rangle $. Consider the case where
the maximal scale is absent ($k_{\min }\rightarrow 0$) and $m_{0}=0$ (or at
least $m_{0}\ll k_{\max })$. Then we can use the expression $N_{k}=1+\left[
k_{\max }/k\right] $ and this gives%
\[
\delta m^{2}=\frac{\lambda }{4}\frac{k_{\max }^{2}}{\left( 2\pi \right) ^{2}}%
\xi \left( 2\right) ,
\]%
where $\xi \left( 2\right) =\sum_{n=1}^{\infty }1/n^{2}$. Thus, we see that
in the non-trivial topology vacuum state, scalar particles acquire the rest
mass of order $m\sim k_{\max }$ and, therefore, in what follows we can
assume $m_{0}>k_{\max }$.

The contribution from a background of scalar particles depends on specific
realization of the generalized statistics for particles. In the simplest
case we can accept expressions (\ref{ef}) which gives for the spectral
density of fluctuations
\begin{equation}
\Phi ^{2}\left( k\right) =k^{2}N_{k}\left( 1+2n_{k}\right) ,  \label{fl}
\end{equation}%
where $n_{k}$ is the standard occupation numbers for scalar particles. The
case $N_{k}=1$ corresponds to the ordinary field theory and the respective
contribution $n$ is known to be very small. Therefore, in what follows we
neglect the corresponding term in (\ref{fl}). Then $N_{k}-1\approx 0$ as $%
k>k_{\max }$ and the integral in (\ref{poten}) can be cut off by $k=k_{\max }
$. It follows that in this integral we can replace $\omega _{k}$ $\simeq
m_{0}$. Thus, we get for the nontrivial topology correction the value
\[
\delta m_{top}^{2}=\frac{\lambda }{2}\frac{\widetilde{n}}{m_{0}},
\]%
where $\widetilde{n}$ is the density of fictitious scalar particles which
accounts for the renormalization of the charge
\[
\widetilde{n}=\frac{1}{2\pi ^{2}}\int_{0}^{k_{\max }}k^{2}\left[ k_{\max }/k%
\right] n_{k}dk.
\]

For estimates we recall that according to (\ref{hp}) the ground state
contains also hidden scalar particles which in the case $k_{\min }=0$ gives
as $k\rightarrow 0$, $n_{k}^{\ast }\sim \left( k_{\max }/k\right) ^{2}$ and $%
n_{k}^{\ast }=0$ as $k>k_{\max }$. In this case the integral (\ref{fp}) is
divergent $\widetilde{n}\rightarrow \infty $ and if for estimates we accept $%
n_{k}\sim \left( k_{\max }/k\right) ^{2-\delta }$ ($\delta \ll 1$) as $%
k<k_{\max }$, then we find that $n\sim k_{\max }^{3}$, while $\widetilde{n}%
\sim k_{\max }^{3}/\delta \gg n$, i.e., $\widetilde{n}$ can take arbitrary
large values as $\delta \rightarrow 0$.

\section{Effective dimension of the Universe}

The growth of the spectral number of fields which takes place in the range
of wave numbers $k_{\min }\left( t\right) <k<k_{\max }\left( t\right) $
leads to the fact that in this range our Universe has to demonstrate
nontrivial geometric properties. Indeed, as it was shown in the previous
sections in the same range of scales a stable equilibrium distribution of
baryons requires a fractal behavior with dimension $D\approx 2$, while the
Newton's and Coulomb's energies of interaction between point particles show
the logarithmic behavior. We recall that the logarithmic potential $\ln
\left( r\right) $ gives the solution of the Poisson equation with a point
source for two dimensions. Both these phenomena are in agreement with the
observed picture of the Universe and it turns out that they have a common
pure geometrical interpretation. Namely, we can say that in the range of
scales $r_{\min }<r<r_{\max }$ (where $r_{\min }=2\pi /k_{\max }$) some kind
of reduction of the dimension of space happens.

Indeed, the simplest way to demonstrate this is to compare the spectral
number of modes in the interval between $k$ and $k+dk$ in MOFT, which is
given by the measure $N_{k}d^{3}k/\left( 2\pi \right) ^{3}$, with the
spectral number of modes for $n$ dimensions in the standard field theory $%
d^{n}k/\left( 2\pi \right) ^{n}$. Hence we can define the effective
dimension $D$ of space as follows
\begin{equation}
k^{3}N_{k}\sim k^{D}.  \label{dim}
\end{equation}%
In the standard field theory $N_{k}=1$ and we get $D=3$, while in MOFT the
properties of the function $N\left( k\right) $ were formed during the
quantum period in the evolution of the Universe and depend on specific
features of topology transformation processes. Thus, in general case, the
effective number of dimensions $D$ may\ take different values for different
intervals of scales. If we take the value (\ref{NN}) we find that in the
range of wave numbers $k_{\max }\geq k\geq k_{\min }$ (where the function $%
N_{k}$ can be approximated by $N_{k}\sim k_{\max }/k$) the effective
dimension of space is indeed $D\approx 2$. The scale $r_{\min }=2\pi
/k_{\max }$ can be called an effective scale of compactification of $3-D$
dimensions, while the scale $r_{\max }$, if it really exists, represents the
boundary after which the dimension $D=3$ restores. We note also that after
this scale the fractal picture of the Universe crosses over to homogeneity
and the standard Newton's law restores.

In MOFT, $N_{k}$ represents the operator of the number of fields ( \ref{N(k)}%
) which is common for all types of fields and, therefore, plays the role of
the density operator for the momentum space. We note that $N_{k}$ is an
ordinary function only in the case when topology changes are suppressed. In
the coordinate representation the number of fields is described by an
operator $N\left( x\right) $ (e.g., see the example of a one-dimensional
crystal) which defines the density of physical space, i.e., the volume
element is given by $dV=N\left( x\right) d^{3}x$.

Consider the relation between these two operators. In what follows we, for
the sake of convenience, consider a box of the length $L$ and will use the
periodic boundary conditions (i.e., $\mathbf{k}=2\pi \mathbf{n}/L$ and, as $%
L\rightarrow \infty $, $\sum_{k}\rightarrow \int \left( L/2\pi \right)
^{3}d^{3}k$). From the dynamical point of view the operator $N_{k}$ has a
canonically conjugated variable $\vartheta \left( k\right) $ such that
\begin{equation}
\left[ \vartheta \left( k\right) ,N_{k^{\prime }}\right] =\vartheta
_{k}N_{k^{\prime }}-N_{k^{\prime }}\vartheta _{k}=i\delta _{k,k^{\prime }}\,.
\end{equation}%
These two operators can be used to define a new set of creation and
annihilation operators
\begin{equation}
\Psi ^{+}\left( k\right) =\sqrt{N_{k}}e^{i\vartheta _{k}},\Psi \left(
k\right) =e^{-i\vartheta _{k}}\sqrt{N_{k}}  \label{Psi}
\end{equation}%
which obey the standard commutation relations
\begin{equation}
\left[ \Psi \left( k\right) ,\Psi ^{+}\left( k^{\prime }\right) \right]
=\delta _{k,k^{\prime }}
\end{equation}%
and have the meaning of the creation/annihilation operators for field modes
(factorized over the number of particles). Thus the density operator for the
momentum space is defined simply as $N_{k}=\Psi ^{+}\left( k\right) \Psi
\left( k\right) $.

In the case when topology transformations are suppressed the operator $\Psi $
can be considered as a classical scalar field $\varphi $ which characterizes
the density of physical space and, in general, depends on time and space
coordinates. We stress that this field has no a direct relation to Higgs
fields discussed in the previous section. In what follows we consider the
spatial dependence only, while the time dependence can be found in the same
way. Indeed, in applying to the ground state $\Phi _{0}=\left\vert
N_{k}\right\rangle $ (which is expressed by the occupation numbers (\ref{d})
and ( \ref{f})) the operators $\Psi $ and $\Psi ^{+}$ change the number of
modes by one, while the total number of modes $N\rightarrow \infty $. In
this sense the state $\Phi _{0}^{\prime }=\Psi \Phi _{0}\approx \Phi _{0}$.
Thus, the classical field $\varphi $ may be defined by relations
\begin{equation}
\varphi \left( k\right) =\left\langle N_{k}-1\left\vert \Psi \left( k\right)
\right\vert N_{k}\right\rangle ,\;\varphi ^{\ast }\left( k\right)
=\left\langle N_{k}\left\vert \Psi ^{+}\left( k\right) \right\vert
N_{k}-1\right\rangle ,  \label{sc}
\end{equation}%
which gives $\varphi \left( k\right) =\varphi ^{\ast }\left( k\right) =\sqrt{%
N_{k}}$. The coordinate dependence of this field can be found in the
standard way (e.g., see Ref. \cite{LL})
\begin{equation}
\Psi \left( \mathbf{r}\right) =e^{-i\mathbf{r}\widehat{\mathbf{P}}}\Psi
\left( 0\right) e^{i\mathbf{r}\widehat{\mathbf{P}}}
\end{equation}%
where $\widehat{\mathbf{P}}$ is the total momentum operator. If in states $%
O_{i}$ and $O_{f}$ the system possesses fixed momenta $\mathbf{P}_{i}$ and $%
\mathbf{P}_{f}$, then
\begin{equation}
\left\langle O_{i}\left\vert \Psi \left( \mathbf{r}\right) \right\vert
O_{f}\right\rangle =\exp \left( -i\mathbf{rk}_{if}\right) \left\langle
O_{i}\left\vert \Psi \left( 0\right) \right\vert O_{f}\right\rangle
\label{x}
\end{equation}%
where $\mathbf{k}_{if}=\mathbf{P}_{i}-\mathbf{P}_{f}$. We note that the
creation/annihilation of a single mode is accompanied with the
increase/decrease in the number of \textquotedblright
dark\textquotedblright\ particles (\ref{hp}) by $N_{k}$ and hence in the
total momentum by $\mathbf{P}_{k}=zN_{k}\mathbf{k}$, where $z$ is the number
of different types of bosonic fields. Thus, from (\ref{x}) and (\ref{sc}) we
find that coordinate dependence will be described by the sum of the type
\begin{equation}
\varphi \left( x\right) \sim \sum_{k}c_{k}\exp \left( -i\mathbf{P}_{k}%
\mathbf{r}\right) .  \label{s}
\end{equation}%
where $c_{k}\sim \sqrt{N_{k}}$. In the range of wave numbers $k_{\max }\geq
k\gg k_{\min }$ we can use the approximation $N_{k}\simeq 1+k_{\max }/k$ and
let the length of the box be $L<r_{\max }=2\pi /k_{\min }$. Then we find
that in the sum (\ref{s}) the momentum $\left\vert \mathbf{P}_{k}\right\vert
\geq zk_{\max }$ and, therefore, the maximal wavelength is indeed restricted
by
\begin{equation}
\ell \leq \ell _{0}<L,
\end{equation}%
where $\ell _{0}=2\pi /zk_{\max }=r_{\min }/z$, which means that at least in
one direction the space box is effectively compactified to the size $\ell
_{0}$. From the physical standpoint such a compactification will be
displayed in irregularities of the function $N\left( x\right) \sim
\left\vert \varphi \left( x\right) \right\vert ^{2}$ (we point out that the
time dependence will randomize phases in (\ref{s})). E.g., the function $%
N\left( x\right) $ may take considerable values $N\left( x\right) \sim <N>$
only on thin two dimensional surfaces of the width $\sim \ell _{0}$ and
rapidly decay outside, which may represent another view on our explanation
of the formation of the observed fractal distribution of galaxies and the
logarithmic behavior of Newton's potential. When considering the box of the
larger and larger size $L$ $\gg r_{\max }$ we find that $N_{k}\rightarrow
N_{\max }\sim r_{\max }/r_{\min }$ and the effect of the compactification
disappears (no restrictions on possible values of wavelengths emerge). Thus
at scales $\ell >r_{\max }$ dimension $D=3$ restores which restores both the
standard Newton's law and the homogeneous distribution of galaxies.

\section{Conclusions}

We have shown that MOFT predicts a rather interesting physics for the range
of scales $r_{\min }<r<$ $r_{\max }$. First of all we point out that in this
range the Universe acquires features of a two-dimensional space whose
distribution in the observed 3-dimensional volume has an irregular
character. This provides a natural explanation to the observed fractal
distribution of galaxies and dark matter, which can be described by the
presence of fictitious particles. Fictitious particles carry charges of all
sorts and may be responsible for the origin of Higgs fields and the
formation of the diffuse X-ray background. We recall that such properties
originate from a primordial thermodynamically equilibrium state and are in
agreement with the homogeneity and isotropy of the Universe. Thus, such a
picture of the Universe represents a homogeneous background, while
gravitational potential fluctuations should be considered in the same way as
in the standard model. Nevertheless, we can state that in the range of
wavelengths $r_{\min }<\lambda <$ $r_{\max }$ the propagation of
perturbations will correspond to the 2-dimensional law.

The fictitious character of particles which compose dark matter leads to
strong correlations between perturbations in the actual baryon density and
perturbations in dark matter. In particular, this also means that in the
range of scales $r_{\min }<r<$ $r_{\max }$ the standard dust-like equation
of state for baryons is invalid and we might expect that this could allow to
explain the so-called dark energy (or quintessence) phenomenon.

We note that Modified Field Theory represents nothing more but the standard
field theory extended to the case where the topology of space is nontrivial
and, in general, may change. In this sense, from the qualitative standpoint
predictions of MOFT are inevitable. If we believe that in the early Universe
topology changes did occur, we merely have to observe now effects related to
dark matter or fictitious particles. In the present paper we have shown that
MOFT is a self-consistent, from particle physics standpoint, theory which
can be readily used in applied problems. Yet, in the present form it is not
exact. It requires\thinspace , in the first place, an explicit description
of topology transformations. There is no formal mathematical problems though
(an appropriate term can be easily added to the Hamiltonian), however an
adequate model for such processes, with a clear physical interpretation, is
still missing. The importance of this problem is based on the observation
that at very small scales ($\ell \leq \ell _{Pl}$) topology transformations
still take place and those should define the behavior of the spectral number
of fields $N_{k}$ in the limit $k\rightarrow \infty $, i.e., the structure
of space and the effective dimension at small scales. It is clear that in
the exact theory ultraviolet divergencies should be absent, which means that
integrals of the type (\ref{poten}) as $r\rightarrow 0$ should have finite
values. This can be fulfilled if the spectral function is restricted by $%
N_{k}<k^{D-3}$ as $k\rightarrow \infty $ with $D<1$ (i.e., at small scales
the effective dimension is less than $1$ or, in other words, the space
disintegrates).

The situation is somewhat better with cosmological production of the field
number density in the very early Universe. In this case there exists at
least a clear qualitative model which can be used to illustrate the process,
e.g. see. Ref. \cite{kir94}. That model is based on the fact that near the
singularity inhomogeneities of the metric have always a large scale
character (i.e., $\ell _{in}\gg \ell _{h}\sim 1/t$, e.g. see, Ref. \cite{K93}%
) and this was shown to be valid during the quantum stage up to the moment
when the classical space emerges (the moment of the origin of the classical
space corresponds to $\ell _{in}\sim \ell _{h}$ \cite{KMq97}). Thus, every
region of space of the size $L\lesssim \ell _{h}$ can be described in the
leading order by a homogeneous mixmaster model. We can say that near the
singularity the Universe splits into a set of homogeneous models, with
parameters of the models varying in space. Then, to describe the topology
formation process we can use the so-called third quantization scheme \cite%
{R88,HM,MG,K92} for every such model. In this scheme the number of
universes $N_{i}\left( x\right) $ in a quantum state $i$ will
correspond to the number of fields in a region $x$ of the size
$L$. In general, the number of universes (or fields) $N_{i}\left(
x\right) $ produced depends essentially on the choice of the
initial quantum state. However, there exists a region of initial
states which gives rise to the thermal distribution with $T\sim
T_{pl}$ (e.g., see Refs. \cite{HM,K92}). Thus, there is always a
room to speculate on the most natural initial conditions, the
measure of initial conditions (and other stupid things which prove
nothing but sound significant).

In the present paper we have supposed that upon the quantum period in the
evolution of the Universe the spectral number of fields $N_{k}$ is
conserved, which means that $N_{k}$ depends on time via only the
cosmological shift of scales (i.e., $k\left( t\right) \sim 1/a\left(
t\right) $, where $a\left( t\right) $ is the scale factor). However, there
are reasonable arguments which show that the number of fields should
eventually decay. Indeed, if we take into account thermal corrections to $%
N_{k}$, then instead of (\ref{NN}) we get the expression of the type $%
N_{k}\sim 1+T_{\ast }/k+...$. In this case the temperature $T_{\ast }$ plays
the role of the maximal wave number $k_{\max }$. Why at present this
temperature is so small $T_{\ast }\ll T_{\gamma }$ (where $T_{\gamma }$ is
the CMB temperature) requires a separate explanation. The situation is
different when we allow the number of modes to decay. Phenomenologically,
the decay can be described by the expression of the type $N_{k}\sim 1+\left(
N_{k}^{0}-1\right) e^{-\Gamma _{k}t}$ (where $\Gamma _{k}$ is the period of
the half-decay which, in general, can depend on wave numbers $k$). In the
simplest case $\Gamma _{k}=const$ and this would lead to an additional
monotonic increase of the minimal scale $r_{\min }\sim a\left( t\right)
e^{\Gamma t}$, while the maximal scale changes according to the cosmological
shift only $r_{\max }\sim a\left( t\right) $. We note that if the ground
state of fields $\Phi _{0}$ contains hidden bosons (\ref{hp}), the decay of
modes will transform them into real particles and, therefore, the decay will
be accompanied by an additional reheating which should change the
temperature of the primordial plasma (which will produce an additional
difference between $T_{\ast }$ and $T_{\gamma }$).

The additional increase in $r_{\min }$ means that values of all interaction
constants and rest masses of particles decrease with time, for at scales $%
\ell \gg r_{\max }$ the renormalized values of interaction constants have
the form $\widetilde{\alpha }$ $\sim $ $\alpha r_{\max }/r_{\min }$ $\sim $ $%
e^{-\Gamma t}$. This would support the Dirac hypothesis \cite{Dir} and
recently observed suspected variation of the fine structure constant at high
red shifts \cite{R11,R11a,R12}.

We also point out to the violation of the Pauli principle for wavelengths $%
\lambda >r_{\min }$ (more than one fermion can have a wavelength $\lambda $
). Such particles are located in the volume $\gtrsim r_{\min }^{3}$ and at
laboratory scales the portion of states violating the statistics is
extremely suppressed $P\lesssim \left( L/r_{\min }\right) ^{3}$, where $L$
is a characteristic spatial scale of a system under measurement (e.g., if as
such a scale $L$ we take the Earth radius, this factor will be still
extremely small $P\sim 10^{-32}$).

\textbf{Acknowledgement}

I am grateful to F. Columbus for the invitation to write a paper for this
volume, and to D. Turaev for useful discussions at all stages of the
research and a partial collaboration. I am especially grateful to J.
Scoville whose simple questions gave me a new look at the subject. I would
also like to beg my pardon of those authors, whose results have direct
relation to this research and were not cited (for I have only accidental
access to the LANL archive and then I am too lazy to search it thoroughly).

\textbf{Warning:} For the last five years, my attempts to find a
support for this research from Russian funds and institutions were
all in vain. E.g., RFFI steadily refuses to support this research.
The "official science" (or at least big guys in the Russian
Academy of Science) seem to have some serious principle
objections: perhaps, the results do not fit the striking,
fascinating, predictive, solving all problems, etc., inflationary
paradigm, or, may be, "...such subtleties had not been considered
yet ...", and "... this is not what we expect and what we want
from quantum gravity..."(Physical Review referees), or at last "
... starting from scales more than a few $Mpc$ the Newton's law is
well-verified ...", and " ...  it is impossible to construct any
reasonable homogeneous and isotropic model for the gravitational
potential which increases as $\ln r$ ..." (JETP Letters). In any
case I feel it is necessary to warn young scientists about the
situation (please, do not take this approach and all the results
too seriously, for all those are not more but rubbish!).

\end{document}